\documentclass[aps, prb, twocolumn, a4paper, superscriptaddress]{revtex4-2}

\usepackage{graphicx} % Include figure files
\usepackage{dcolumn} % Align table columns on decimal point
\usepackage{amsmath, bm} % maths packages
\usepackage{hyperref} % add hypertext capabilities
\usepackage{mathrsfs} 
\usepackage{color} % colour in document
\usepackage{gensymb} % degree symbol
\usepackage{braket}
\usepackage[utf8]{inputenc}
\usepackage{tikz}
\usepackage{lipsum}
\usepackage{makecell}
\usepackage{multirow}

\makeatletter
\g@addto@macro\bfseries{\boldmath}
\makeatother

\def\cmg{C\lowercase{e}M\lowercase{n}$_2$G\lowercase{e}$_2$}
\def\pmg{P\lowercase{r}M\lowercase{n}$_2$G\lowercase{e}$_2$}
\def\nmg{N\lowercase{d}M\lowercase{n}$_2$G\lowercase{e}$_2$}
\def\remg{$R$Mn$_2$Ge$_2$}
\def\remgall{$R$Mn$_2$Ge$_2$ ($R$=Ce,~Pr,~Nd)}
\def\musr{$\mu$SR}

\begin{document}

\title{Local magnetic properties of the rare-earth intermetallics~\remgall}

\author{N.~P.~Bentley}
\affiliation{Department of Physics, Centre for Materials Physics, Durham University, Durham, DH1 3LE, United Kingdom}

\author{T.~L.~Breeze}
\affiliation{Department of Physics, Centre for Materials Physics, Durham University, Durham, DH1 3LE, United Kingdom}

\author{A.~Hern{\'a}ndez-Meli{\'a}n}
\affiliation{Department of Physics, Centre for Materials Physics, Durham University, Durham, DH1 3LE, United Kingdom}

\author{M.~J.~Pearce}
\affiliation{Department of Physics, Centre for Materials Physics, Durham University, Durham, DH1 3LE, United Kingdom}

\author{T.~J.~Hicken}
\affiliation{PSI Center for Neutron and Muon Sciences, 5232 Villigen PSI, Switzerland}

\author{M.~T.~F.~Telling} 
\affiliation{ISIS Neutron and Muon Source, STFC Rutherford Appleton Laboratory, Harwell, Didcot OX11 0QX, United Kingdom}

\author{B.~M.~Huddart}
%\affiliation{Department of Physics, Centre for Materials Physics, Durham University, Durham, DH1 3LE, United Kingdom}
\affiliation{Clarendon Laboratory, University of Oxford, Department of Physics, Oxford OX1 3PU, United Kingdom}

\author{G.~D.~A.~Wood}
\altaffiliation[Current address: ]{ISIS Neutron and Muon Source, STFC Rutherford Appleton Laboratory, Harwell, Didcot OX11 0QX, United Kingdom}
\affiliation{Department of Physics, University of Warwick, Coventry, CV4 7AL, United Kingdom}

\author{D.~A.~Mayoh}
\affiliation{Department of Physics, University of Warwick, Coventry, CV4 7AL, United Kingdom}

\author{G.~Balakrishnan}
\affiliation{Department of Physics, University of Warwick, Coventry, CV4 7AL, United Kingdom}

\author{S.~J.~Clark}
\affiliation{Department of Physics, Centre for Materials Physics, Durham University, Durham, DH1 3LE, United Kingdom}

\author{T.~Lancaster}
\affiliation{Department of Physics, Centre for Materials Physics, Durham University, Durham, DH1 3LE, United Kingdom}

\begin{abstract}
We present an investigation of the rare-earth intermetallic materials,~\remgall, reported to host a lattice of skyrmionic bubbles at room temperature. The magnetism of all three materials is characterised by the onset of local fluctuations as the magnetic state changes from a conical magnetic structure to a collinear antiferromagnetic one as temperature $T$ is increased. 
In this $T$ regime, where skyrmion-bubble textures have been reported, we see dynamics similar to those observed in other skyrmion-hosting materials. The level of disorder above this transition increases across the series from $R$=Ce to Nd. 
At low temperatures the magnetism is affected by the ordering of the rare-earth ions, resulting in distinct behaviour for the different members of the series.

\end{abstract}

\maketitle

\section{Introduction}
\label{sec:Intro}
Materials that host skyrmions and related excitations show promise for applications in spintronics and quantum computing~\cite{lancaster2019skyrmions,Jiang2015bubbles,fert2017_skyrmions,psaroudaki2023_skyrmionqubits}.
These topological objects have been observed in rare-earth-based intermetallics, which provide the wide range of competing magnetic interactions and short-range geometric frustration needed for skyrmion stabilisation in centrosymmetric systems~\cite{Yu2012helicity,Chakrabartty2022tunsk,bouaziz2022centro,Paddison2022Gd,gomilsek2025muon,Moody2025centrodmi,viviane2025centro}.
The $RM_2X_2$ family of compounds (where $R$ is the rare-earth, $M$ is a transition metal and $X$ represents Si, Ge or As) form a subset of the rare-earth intermetallics that crystallize in a ThCr$_2$Si$_2$-type tetragonal structure (space group $I4/mmm$). %, which consist of stacked layers of ions in the centrosymmetric sequence $M$-$X$-$R$-$X$-$M$. 
Along with topological magnetic excitations, these materials host a wide range of exotic magnetic phenomena, including a variety of multi-q magnetic states~\cite{Wood2023doubleq,Paddison2024Gddynamics,wood2025magnon,Huddart2025Gdmuon,matsuyama2023_GdquantumOsc,dong2025_Gdfermiarc,hayashi2024_GdOs2Si2}, multistep transitions~\cite{Yoshimochi2024_multisteptopology,Littlehale2024_SDW}, unconventional superconductivity~\cite{stockert2011_CeCu2Si2,Yamashita2017_CeCu2Si2}, and competing spin and charge density wave ordering~\cite{moya2022_EuCDW,vibhakar2023_Eucompeting,agarwal2025_EuCDW}.

A particularly interesting case is given by the~\remgall~group of materials, in which magnetic ordering of both the Mn and rare-earth ions results in a rich cascade of magnetic transitions~\cite{welter1995neutrons,MdDin2015Cetunable,Wang2014PrNeutrons,xu2023Ndtopological}.  
These materials are also reported to host skyrmion bubbles (SkBs) (objects that have a similar magnetic structure to the skyrmion, but are not stabilised by the Dzyaloshinskii-Moriya (DM) interaction~\cite{Nagaosa2013skyrmtopology,Gobel2021beyondskyrm,treves2025Ndstability}) and other topological excitations over a wide temperature range, including at room temperature~\cite{hou2021skyrmions,li2024CeSkB}.
Lorentz transmission electron microscopy (LTEM) has been used to image these excitations and, for~\nmg, to construct a phase diagram highlighting the density and type of excitation~\cite{hou2021skyrmions}.  
The anomalous and topological Hall effects are also reported in all these materials~\cite{xu2022CeHall,huang2023Pranisotropy,Wang2023PrHall,Lyu2025_PrHall,zheng2021Ndtopologicalhall,wang2020Ndskyrmions}, suggesting the presence of topological magnetic objects with non-zero chirality.

Here we utilise muon-spin spectroscopy (\musr) and density functional theory (DFT) calculations to investigate the magnetism of~\remgall.
The muon, as a probe of local magnetism~\cite{muontextbook2022}, is sensitive to the many changes of long-range magnetic order reported in~\remgall. 
By modelling the field distributions at candidate muon sites for each of these magnetic structures, we identify a crystallographically unique muon site that captures the changes in local magnetism for all the materials in the series.
In particular, the variation in the low-temperature magnetic behaviour, due to the ordering of the rare-earth ions, has been investigated in detail revealing its evolution across the series. 
\musr~is also sensitive to spin dynamics in~\remgall~which manifest via magnetic fluctuations on the muon timescale~\cite{hicken2021megahertz,gomilsek2025muon,franke2018magnetic,hicken2020magnetism}. 
This sensitivity allows us to identify an onset of dynamics close to the phase transition out of a conical magnetic state as a characteristic feature of each of the materials in this series. The onset occurs in the temperature regime where the SkBs have been identified and we suggest that the dynamics reflect the presence of these excitations. 

The paper is structured as follows: in Sec.~\ref{sec:ExpCompDetails} we describe our experimental and computational methods; in Secs.~\ref{sec:Ce},~\ref{sec:Pr}~and~\ref{sec:Nd} we present results on each of~\remgall, before comparing the behaviour of the compounds in Sec.~\ref{sec:Discussion} and presenting our conclusions in Sec.~\ref{sec:Conclusion}.

\section{Experimental and Computational details}
\label{sec:ExpCompDetails}
\remgall~single crystals were prepared using the in-flux method. High-purity elemental Nd (3N, Strem), Ce (3N, Strem), Pr (3N, Strem), Mn (3N, Alfa Aesar), Ge (6N, ABCR), and In (3N, ABCR) were placed into an alumina Canfield crucible set with the following ratio: R:Mn:Ge:In=1:2:2:30. 
The crucible set was then sealed in a quartz tube under vacuum. The sealed tube was placed into a furnace, heated to $1100~^\circ$C, and held at this temperature for $24$~h. The tube was then slowly cooled to $700~^\circ$C at a rate of $3~^\circ$C h$^{-1}$. At $700~^\circ$C, the tube was removed from the furnace and centrifuged to remove the excess In.
The crystals were initially characterised using energy-dispersive X-ray spectroscopy to confirm their compositions, and magnetometry to confirm the temperature dependence of the magnetic susceptibility.

Zero field (ZF) and longitudinal field (LF)~\musr~measurement of single crystal mosaics of~\remgall~were carried out making use of the GPS and FLAME instruments~\cite{amato2017_GPS} at the Swiss Muon Source (S$\mu$S), Paul Scherrer Institut, with additional measurements performed on~\nmg~using the HIFI instrument~\cite{lord2011hifi} at the STFC-ISIS facility.
The mosaics (each covering an area of 1~cm$^2$) comprising  $\approx10$ crystals were aligned in an Ag foil envelope (foil thickness 25$\mu$m) with their $c$ axis oriented parallel to the muon beam direction. Samples were mounted in fly-past geometry and were inserted in a $^4$He flow cryostat or a closed-cycle refrigerator (CCR) to access temperatures above $300$~K.

In a~\musr~experiment spin polarized positive muons are implanted in the sample and interact with the local magnetic field. 
These muons decay with an average lifetime of $2.2~\mu$s and emit a positron preferentially along the muon spin direction.  
We measure the asymmetry,
\begin{equation}
    A(t)=\frac{N_\mathrm{B}(t)-\alpha N_\mathrm{F}(t)}{N_\mathrm{B}(t)+\alpha N_\mathrm{F}(t)},
    \label{eq:asymmetry}
\end{equation}
where $N_\mathrm{F/B}$ is the number of positron counts in the forward and backwards detectors and $\alpha$ is a correction factor accounting for detector geometry. The asymmetry is proportional to the average spin polarization of the muon ensemble.
In all the measurements the initial muon spin direction was directed antiparallel to the $c$ direction of the mosaics. All measurements made in LF (with a magnetic field also directed along the $c$ axis) employed the set of cooling procedures detailed in Ref.~\cite{hou2021skyrmions} in order to stabilise SkBs and measure their impact on the dynamic response.
Analysis of the muon data was performed using the WiMDA program~\cite{Pratt2000wimda}. 

Calculations of candidate muons sites in~\remgall~made use of the plane-wave, pseudopotential code, \textsc{Castep}~\cite{clark2005first} and the~\textsc{Mufinder}~program~\cite{huddart2021mufinder}. 
For~\cmg~and~\pmg~we generated a set of muonated supercells, consisting of a $2\times2\times1$ cells of~\remg~(with lattice parameters consistent with the experimentally measured values)~and a muon (modelled by an ultrasoft hydrogen pseudopotential) constrained to be at least $1$~\AA~from any atoms and $0.5$~\AA~from any symmetry-related positions occupied by muons in other simulated supercells. 
Twenty supercells of each material were allowed to relax, identifying energetically favourable locations to be occupied by a muon. Each calculation was performed using the PBE functional, a plane-wave cutoff of $1200$~eV and a $2\times2\times2$ Monkhorst-Pack grid~\cite{monkhorst1976special} for Brillouin-zone sampling, giving total energies converged to within $1$ meV per atom.
For~\nmg~we seeded simulation cells with the initial muon sites identified in the other materials.
Geometry optimisation calculations were then performed on this compound, using a plane-wave cutoff of $1200$~eV and $3\times3\times3$ Monkhorst-Pack grid to satisfy the same convergence criteria as above.

Each of~\remgall~are reported to display incommensurate magnetic ordering of the Mn ions in a conical helix~\cite{welter1995neutrons}. The magnetic moment $\mathbf{m}_\mathrm{Mn}$ at position $\mathbf{r} = (x,y,z)$ is given by
\begin{equation}
    \mathbf{m}_\mathrm{Mn}(\mathbf{r})= 
    \begin{bmatrix}
    m_q\cos(2\pi q_zz) \\
    -m_q\sin(2\pi q_zz)\\
    0
    \end{bmatrix}
    +
        \begin{bmatrix}
    0 \\
    0\\
    m_c
    \end{bmatrix},
    \label{eq:conical_mag}
\end{equation}
which generates a conical helix aligned along $c$ with propagation vector $(0,~0,~q_z)$. 
This structure is often parametrised in terms of  the magnetic moment size $\left( m_\mathrm{Mn}=\sqrt{m_q^2+m_c^2}\right)$, cone semi-angle ($\tan\alpha=m_q/m_c$) and propagation vector $(0,~0,~q_z)$.
The resulting dipole field $\textbf{B}_\mathrm{dip}$ at the muon site with position $\mathbf{r}_\mu$ is given by 
\begin{equation}
\label{eq:dipole_field}
    \mathbf{B}_\mathrm{dip}(\mathbf{r}_\mu) = \frac{\mu_0}{4\pi}\sum_{i} \left [\frac{(\mathbf{m}_i\cdot\mathbf{r}_{i})\mathbf{r}_i}{r_i^5} - \frac{\mathbf{m}_i}{r_i^3}
    \right],
\end{equation}
where $\mathbf{r}_i$ is the distance from the muon to the $i^{th}$ magnetic ion in the materials and $\mathbf{m}_i$ is the magnetic moment of the magnetic ion at position $\mathbf{r}=\mathbf{r}_i+\mathbf{r}_\mu$.
This sum was evaluated using the MuESR program~\cite{bonfa2018introduction}, using the method described in Ref.~\cite{martin2016MnGe} for incommensurate helices.

In general, helical spin arrangements result in an Overhauser-type field distribution whose average sits at a non-zero value of field. The resulting frequency spectrum will be the Fourier transform of this, involving the multiplication of an oscillation at a frequency corresponding to the average field, and the Fourier transform of an Overhauser distribution, which is a Bessel function with a characteristic frequency corresponding to the difference in field between the sharp peaks and the average field~\cite{muontextbook2022}. This frequency spectrum is often well approximated by two frequencies corresponding to the two sharp peaks in the field distribution, or a single peak at the average field, in the case that the distribution is broadened. 

\section{\cmg}
\label{sec:Ce}
\cmg~is believed to host the simplest set of magnetic transitions of the~\remgall~family, due to the reported absence of rare-earth magnetic ordering at low temperatures~\cite{welter1995neutrons,FernandezBaca1996_Ceneutrons,MdDin2015Cetunable}. 
The magnetic states reported comprise the following phases:
\begin{enumerate}
\itemsep-0.25em 
\item[(a)] I$_c$ phase ($T<318$~K), where Mn spins form an incommensurate conical structure with a ferromagnetic (FM) component aligned along the $c$ axis and a helical component in the $a$-$b$ plane. 
\item[(b)] AFM phase ($318\leq T \leq 417$~K), with Mn spins in a collinear antiferromagnetic (AFM) structure within the $a$-$b$ plane.
\item[(c)] PM phase ($T>417$~K), with disordered Mn spins. 
\end{enumerate}
The transitions between these phases are denoted as $T_c=318$~K and $T_\mathrm{N}=417$~K.

\subsection{\musr~results}
\begin{figure*}
  \centering
  \includegraphics[width=0.9\linewidth]{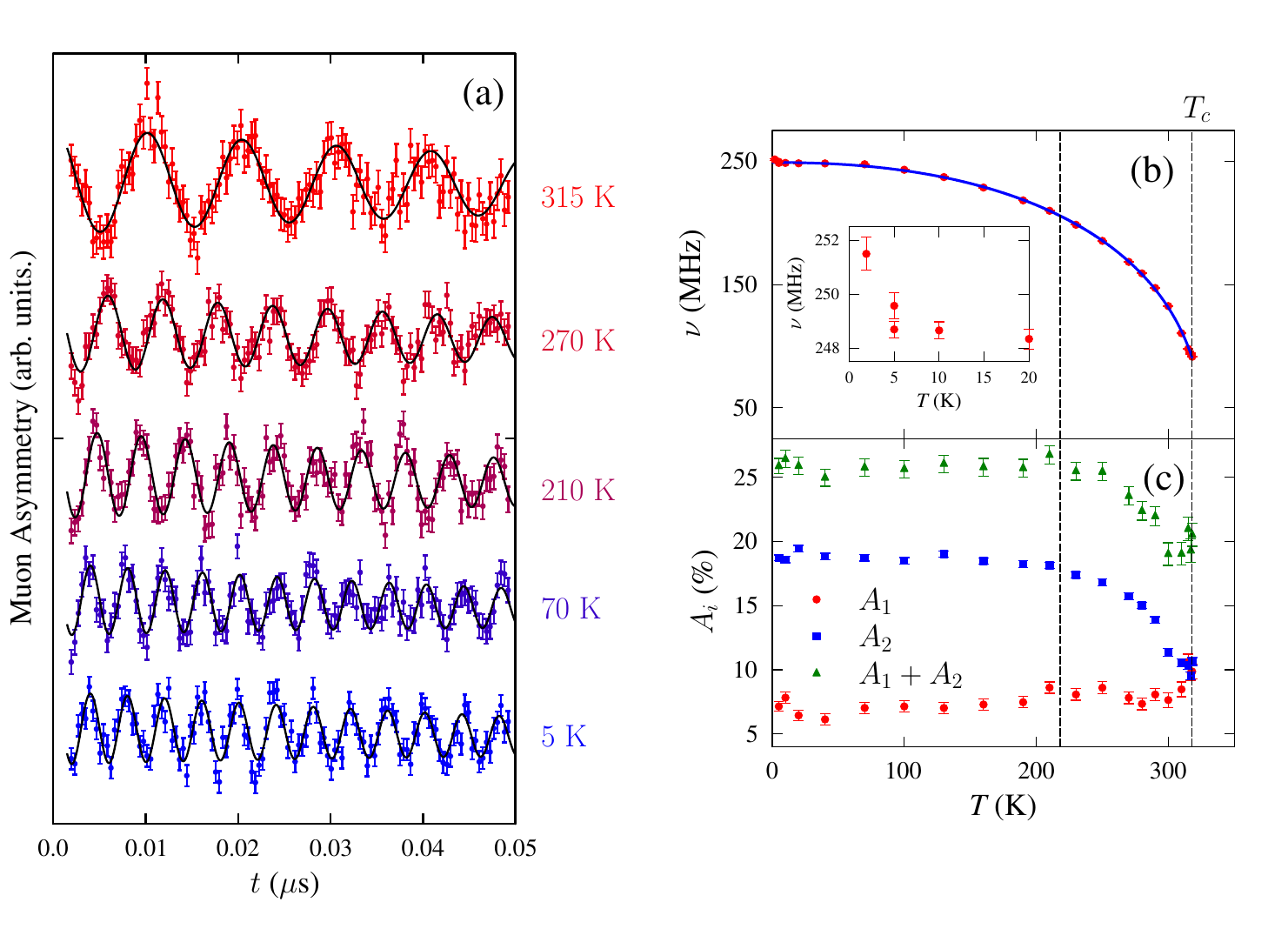}
  \caption{ZF~\musr~measurements of~\cmg~demonstrating rapid oscillatory behaviour in the first $0.05~\mu$s of the time window. (a) Asymmetry spectra below $T_c=318$~K, with the change in (b) precession frequency and (c) amplitudes, $A_1$ (red circles), $A_2$ (blue squares), and $A_1+A_2$ (green triangles) with temperature. The dotted line at $218$~K marks the onset of  changes in the amplitudes due to the spin reorientation transition at $T_{c}$.} 
  \label{fig:Ce_PSI_ZFsmallt}
\end{figure*}
\musr~reveals oscillations below $T_c=318$~K, reaffirming reports of magnetic ordering in~\cmg. The rapid oscillatory behaviour in the first 0.05~$\mu$s of the ZF~\musr~spectra [Fig.~\ref{fig:Ce_PSI_ZFsmallt}(a)] is captured by 
\begin{equation}
	A\left(t\right) = A_{1}e^{-\lambda_1t}\cos(2\pi\nu t) + A_{2}e^{-\lambda_2t}, 
    \label{eq:Ce_osc}
\end{equation}
 where $\lambda_1=12(1)~\rm\mu s^{-1}$ and $\lambda_2=1.9(2)~\rm\mu s^{-1}$ are globally refined and held constant.
The first term in Eq.~\ref{eq:Ce_osc} accounts for the component of muon spin that is initially perpendicular to the internal field, resulting in a rapid precession of the spin, while the second term accounts for components of the muon spin initially parallel to the internal field direction. The parallel component can only be relaxed by dynamic fluctuations in the local field, leading to relatively slow relaxation (with relaxation rate $\lambda_2$).

Order-parameter-like behaviour of frequency $\nu(T)$ is apparent below $T_c$ in Fig.~\ref{fig:Ce_PSI_ZFsmallt}(b), suggesting the ordering of a single species of magnetic ion down to at least $5$~K (ordering of a second ion species  might be expected to change how the local field at the muon site varies with temperature).
However, a discontinuous transition in frequency occurs at $318$~K, above which oscillations are no longer observed.
The amplitude of both components in Eq.~\ref{eq:Ce_osc} is constant in the range $0 \lesssim T \lesssim 218$~K
[Fig.~\ref{fig:Ce_PSI_ZFsmallt}(c)], above which 
a gradual decrease in $A_2$ and total amplitude, $A_1+A_2$, is seen. 
Above $300$~K this behaviour is accompanied by a small increase in $A_1$, corresponding to the larger amplitude of oscillation seen in the spectrum at $315$~K in Fig.~\ref{fig:Ce_PSI_ZFsmallt}(a) compared to lower temperatures.
The temperature at which the amplitudes start to change ($218$~K) coincides with the onset of a reported reorientation of the Mn spins into the $a$-$b$ plane~\cite{MdDin2015Cetunable,xu2022CeHall}. 

At temperatures less than $5$~K a small increase of $\approx2$~MHz is seen in the precession frequency [Fig.~\ref{fig:Ce_PSI_ZFsmallt}(b) inset].
We note that this behaviour is similar to a change we observe at low temperatures in~\nmg, (see Section~\ref{sec:Nd}), which, in that case, corresponds to the magnetic ordering of the Nd ions~\cite{welter1995neutrons,xu2023Ndtopological}.
Therefore we suggest it is likely that the Ce ions order at low temperatures; this is also supported by our dipole field calculations described below.

\begin{figure}
  \centering
  \includegraphics[width=0.9\linewidth]{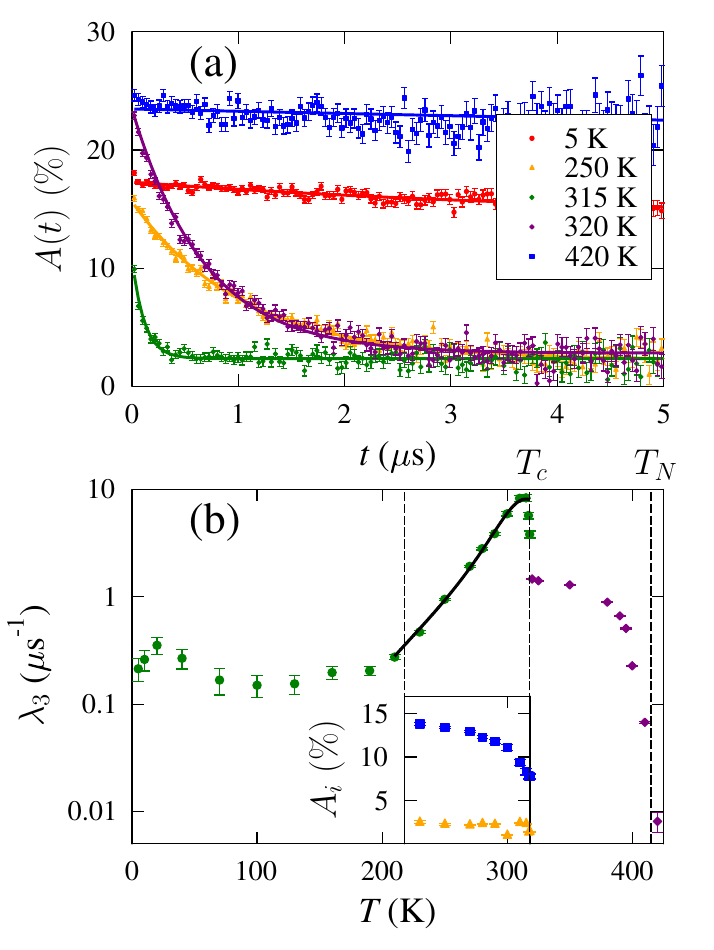}
  \caption{ZF~\musr~measurements of~\cmg~demonstrating relaxing behaviour over $5~\mu$s. (a) Asymmetry spectra at several temperatures, showing the missing asymmetry associated to the oscillating fraction being recovered above $T_c$. (b) Change in relaxation rate with temperature [with an inset showing the decrease in relaxing amplitude $A_3$ (blue squares)], that occurs between $218$~K and $T_c$ (green circles). The fit of Eq.~\ref{eq:Lorentzian_slowrelaxation} to $\lambda_3$ is shown in black and the background contribution, $A_4$ (orange triangles), is approximately constant. Above $T_c=318$~K both amplitudes are globally refined and held constant, and $\lambda_3$ points are plotted as purple diamonds.}
  \label{fig:Ce_PSI_ZFlarget}
\end{figure}

Above $T_c=318$~K no oscillations are observed, despite the evidence from other experimental techniques that~\cmg~remains magnetically ordered as an AFM within the $a$-$b$ plane up to $417$~K~\cite{welter1995neutrons,MdDin2015Cetunable}. 
In this temperature region it is instructive to look at the slow relaxation of the muon-spin ensemble (in contrast to the response within the first 0.05~$\mu$s where the oscillations were resolved). In order to do this, the data were arranged in time bins in which the oscillatory signal, which has $\nu>50$~MHz, is unresolvable, but the relaxation over several $\mu$s can be reliably fitted.
The resulting magnetic response of muons over an $8~\mu$s time window at all measured temperatures is described by
\begin{equation}
	A\left(t\right) = A_{3}e^{-\lambda_3t} + A_{4},
    \label{eq:Ce_rel}
\end{equation}
where $A_{3}=20.7(1)\%$ and $A_{4}=2.8(1)\%$ are globally refined and held constant above $T_c=318$~K. 
The term with amplitude $A_3$ captures the magnetic dynamics and $A_4$ provides a constant term describing muons implanting in non-magnetic environments, including the silver foil packet and the cryostat walls.
Comparing to the spectrum above $T_\mathrm{N}=417$~K, a missing initial asymmetry fraction is evident below $T_c$ in Fig.~\ref{fig:Ce_PSI_ZFlarget}(a),  corresponding to the oscillating amplitude, $A_1$. This fraction increases as the transition is approached from below, reflecting the increase of $A_1$ close to $T_c$ seen in Fig.~\ref{fig:Ce_PSI_ZFsmallt}(c). 
All of the initial asymmetry is recovered above $T_c$ at $T=320$~K, with a large relaxing fraction persisting below $T_\mathrm{N}$.

Of particular interest is the behaviour of the longitudinal relaxation rate $\lambda_3$ [Fig.~\ref{fig:Ce_PSI_ZFlarget}(b)], which reflects dynamic fluctuations in the local magnetism. 
A small peak in $\lambda_3$ is seen at low temperatures ($T<30$~K, see below), above which the relaxation rate is approximately constant until $200$~K. 
Between $200$~K and just below $T_c=318$~K, $\lambda_3$ increases rapidly, and roughly exponentially, with temperature, before peaking around $315$~K, which is below the spin reordering transition at $T_{c}=318$~K. 
Similar behaviour has been seen in several skyrmion-candidate materials in applied longitudinal field~\cite{hicken2020magnetism,hicken2021megahertz,wilson2021_spindynamics} and was attributed to a temperature-dependent reduction in the characteristic fluctuation rate $\tilde{\nu}$ of the dominant relaxation process as a transition is approached from below, such that, at the peak,  $\tilde{\nu}$ is in resonance with the muon precession frequency $\frac{\gamma_{\mu}}{2\pi} B$, where $B$ is the magnetic field at the muon site. 

For~\cmg~in ZF, the static field at the muon site is determined by the Mn magnetic structure and any polarization of the Ce ions. 
Motivated by the peaked behaviour in $\lambda_{3}$ in this, and the other members of the series described below, we parametrize the low-temperature part of the peak with the Lorentzian function 
\begin{equation}
    \lambda=\frac{A}{1+\left[ (T-T_\mathrm{peak})/\Gamma\right]^2 }+ \lambda_0,
    \label{eq:Lorentzian_slowrelaxation}
\end{equation}
where $T_\mathrm{peak}=315$~K corresponds to the temperature of the peak in the relaxation rate.
This gives a parametrization of the width of $\Gamma=26(1)$~K, corresponding to a characteristic energy of $2.2$~meV. 
The peak in $\lambda_{3}$ is accompanied by a decrease in the relaxing amplitude [Fig.~\ref{fig:Ce_PSI_ZFlarget}(b) inset] that was also seen in the first $0.05~\mu$s of the data, and accompanies the spin reorientation of moments into the $a$-$b$ plane as the AFM state is realised above $T_{c}=318$~K.

\subsection{Muon site analysis}

In order to interpret our results in each of the members of this series, we computed candidate muon sites. 
The application of the DFT+$\mu$ method~\cite{blundell2023_DFTmu} using $20$ inital sites in~\cmg~allows us to identify four classes of candidate muon site. 
We find that the magnetic behaviour in~\cmg~is best captured by a site  at (0, 0, 0.2). This site is also found to best describe the measurements in the other members of the series. Subsequent muon site analysis on each material focusses on this site, with a comparison the other candidate muon sites presented in the Supplementary Material~\cite{sm_muon}.

\begin{figure*}
  \centering
  \includegraphics[width=1.0\linewidth]{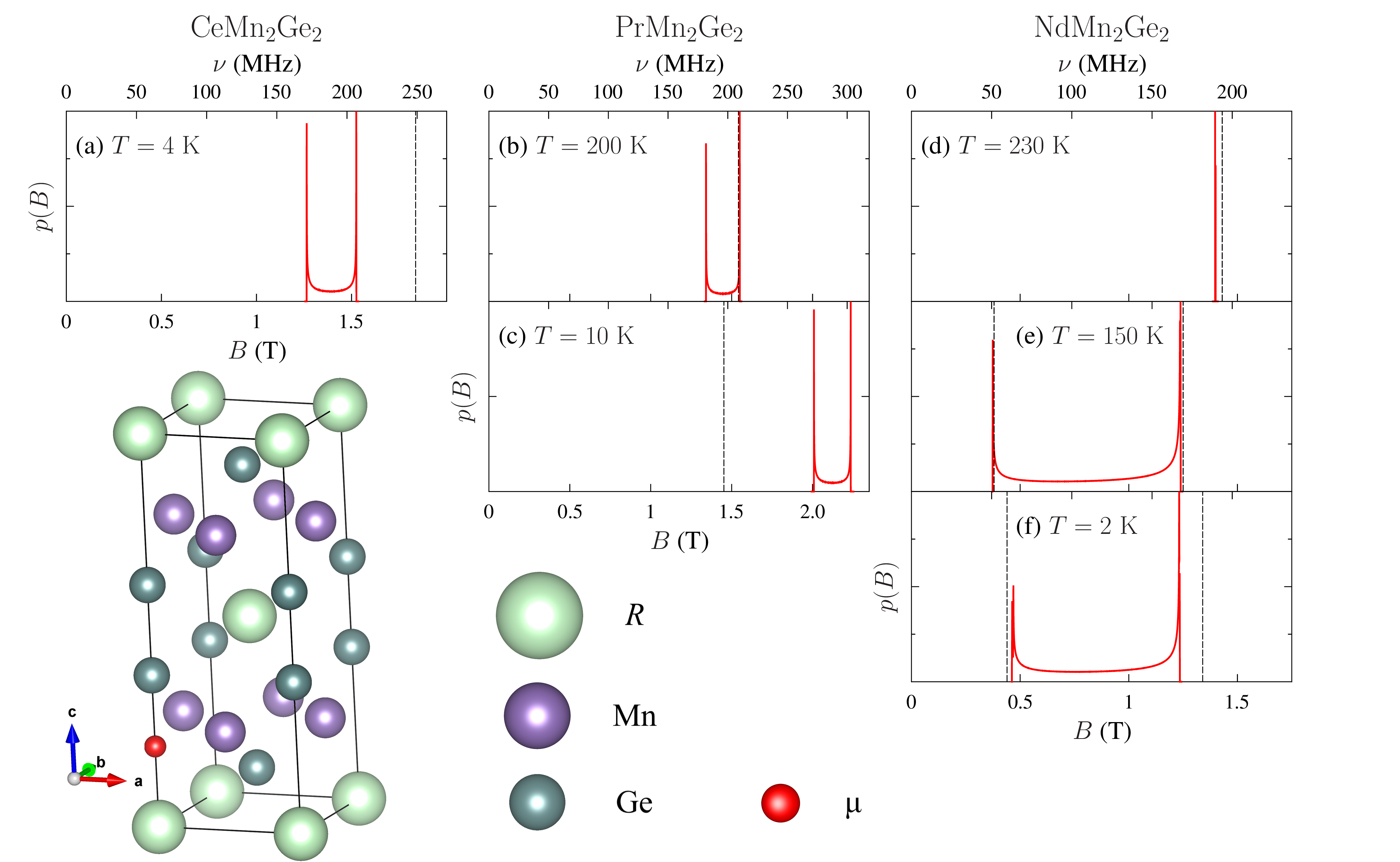}
  \caption{Dipole field distributions calculated at the muon site for reported incommensurate magnetic orderings in~\remgall, with the muon site shown in the structure of~\remgall. In~\cmg, the field distribution corresponding to the measured helical magnetic structure~\cite{MdDin2015Cetunable} at (a) $4$~K is shown, with the dotted line corresponding to the precession frequency of $249.0(5)$~MHz from our~\musr~measurements. 
  In~\pmg~the field distributions for the reported magnetic structures~\cite{Wang2014PrNeutrons} at (b) $200$~K and (c) $10$ are presented. The dotted lines indicate the value of the local field from our~\musr~measurements, which corresponds to precession frequencies of $209.0(4)$~MHz at $200$~K and $196.8(4)$~MHz at $10$~K.
  In~\nmg~the field distributions for the reported magnetic structures~\cite{welter1995neutrons,xu2023Ndtopological} at (d) $230$~K, (e) $150$~K, where the helical axis has reorientated along $b$, and at (f) $2$~K, where both the Mn and Nd moments are ordered, are displayed. The dotted lines indicate the value of the local field from our~\musr~measurements, which corresponds to precession frequencies of $193.9(5)$~MHz at $230$~K, $170.0(8)$~MHz and $51.4(8)$~MHz at $150$~K, and $181.8(7)$~MHz and $60.1(7)$~MHz at $2$~K.}
  \label{fig:R_field_dist}
\end{figure*}

The conical magnetic structure of Mn ions reported at $4$~K, revealed by neutron scattering measurements on~\cmg~\cite{welter1995neutrons,MdDin2015Cetunable}, is characterised by $m_\mathrm{Mn}=3.17\mu_\mathrm{B}$, $\alpha=53^\circ$ and $q_z=0.317$, with the helical component of the Mn spins AFM aligned within the $a$-$b$ plane. The computed local field distribution at the muon site, shown in Fig.~\ref{fig:R_field_dist}(a), gives a range of frequencies that are somewhat smaller than the~\musr~measurement frequency at 4~K. 
Our dipole field calculations do not include the hyperfine contribution to the local field at the muon site due to the local spin density~\cite{onuorah2018hf},
which is difficult to calculate in the case of incommensurate magnetic ordering.  We note that a hyperfine field of magnitude $440$~mT would account for the discrepancy in the average local field at the muon site, and is a reasonable value for the contact field in~\cmg~when compared to values quoted for similar materials~\cite{Campbell1984_rareearthcontact,Schreier2000_metallicREcontact,Ishant2025_Cecontact,Huddart2025Gdmuon}.  

In the AFM phase above $318$~K, where the Mn spins alternate orientation within the $a$-$b$ plane and along the $c$ axis, the Mn moment size is reported as $\approx2.0\mu_\mathrm{B}$ at $350$~K~\cite{MdDin2015Cetunable}.  
As the direction of the Mn spins in the $a$-$b$ plane is unknown, we calculated the local field for a number of possible orientations.
We find that a significant component of the local field lies within the $a$-$b$ plane for all spin orientations. This implies that the observed absence of muon-spin oscillations in this phase is not due simply to the orientation of the local field along the initial muon spin direction, leading to the observation of only the dynamical relaxation of the muon spin. 

To investigate whether rare earth ordering accounts for the small increase in precession frequency seen at low temperature in the inset of Fig.~\ref{fig:Ce_PSI_ZFsmallt}(b), we  simulated the effect of Ce ordering on the local field at the muon site. A FM ordering of Ce spins parallel to the FM component of the Mn spins (along $c$), as seen in the other~\remg~materials, leads to a reduction in the average local field at the muon site. 
In contrast, the ordering of the Ce spins along $-c$ (anti-parallel to the FM component of the Mn spins) leads to an increase in the local field. 
To match the experimentally observed increase in precession frequency of $2.8$~MHz at low temperatures, a Ce moment of $m_\mathrm{Ce}\approx0.4\mu_\mathrm{B}$ was required. 

Although no evidence of Ce order has been seen with neutron scattering~\cite{welter1995neutrons,MdDin2015Cetunable}, time differential perturbed $\gamma$-$\gamma$ angular correlation (TDPAC) spectroscopy~\cite{Frauenfelder1965_TDPAC,Webb2013_TDPAC} identifies a sharp deviation from the expected behaviour of the magnetic hyperfine field at low temperatures~\cite{Carbonari2004Cehyperfine}. 
This was initially explained as a local polarisation of the Ce ions due to the ordered Mn moments, but further analysis~\cite{Lalic2004_CelowTorder} suggests that the Ce moments are ordered, but are very small due to the near cancellation of spin and orbital contributions. 
The estimated Ce moment in this picture is $\approx0.16\mu_\mathrm{B}$ antiparallel to the Mn moments.
The difference between this and the moment value we suggest based on the change in oscillation frequency ($0.4\mu_\mathrm{B}$) could be due to a change in the hyperfine contribution to the local field at the muon site, due to Ce ordering changing the electron spin density at the muon site.

\subsection{Discussion}
The absence of oscillations above $T_c=318$~K 
is unexpected since the material remains magnetically ordered up to $T_{\mathrm{N}}=417$~K. 
However, this lack of oscillations is likely due to the nature of the dynamic fluctuations of the Mn magnetism in the AFM phase.
The observation of muon-spin precession oscillations in a magnetically ordered state requires that (i) there is a significant component of the ordered magnetic field perpendicular to the initial muon-spin direction; and (ii) that the width of the distribution of disordered fluctuating fields $\Delta=\gamma_\mu\sqrt{\langle B^2\rangle - \langle B\rangle^{2}}$ is much less than the average field at the muon site $\langle B\rangle$, i.e.\ $\Delta \ll \langle B\rangle$. Our dipole field simulations suggest that condition (i) is satisfied in this phase. 

To address (ii), we note that magnetic fluctuations in a material can be characterised by a rate $\tilde{\nu}=1/\tau$ (where $\tau$ is the fluctuation time), and an amplitude expected to be of the order of $\Delta$ defined above.
In the limit of fast fluctuations, we have $\Delta\tau\ll 1$ and dynamics cause an exponential decay of the asymmetry at a rate $\lambda_3=2\Delta^2\tau$~\cite{muontextbook2022} in ZF.
Therefore, the observation of exponential relaxation [see Fig.~\ref{fig:Ce_PSI_ZFlarget}(a)] is consistent with dynamic relaxation in the fast-fluctuation limit. 
This suggests it is likely that the muon ensemble experiences a range of local fields satisfying $\Delta \gtrsim \langle B\rangle $, with fluctuation time satisfying $\Delta \tau\ll 1$.

It is notable that magnetic Bragg peaks are observed in this temperature regime~\cite{MdDin2015Cetunable}. Compared to the MHz timescale of~\musr, neutrons effectively take a snapshot of the magnetism~\cite{boothroyd2020_textbook}, which evidently is ordered over a time scale $\tau> 10^{-11}$~s to allow magnetic Bragg peaks to be observed. Although it is harder to make a quantitative estimate of correlation lengths, we note that $\Delta$ increases such that oscillations tend not to be resolved if magnetically ordered regions contract below around 10 lattice spacings.
On the basis of our measurements, we suggest that $\Delta/\gamma_\mu\gtrsim 2$~T and therefore $\tau \lesssim 5\times10^{-10}$~s. 
If $\tau$ is only weakly $T$-dependent in this regime, then the relaxation rate will vary as $\lambda_{3} \sim  \Delta^{2}$ (proportional to the square of the magnitudes of the local fields),  justifying the order parameter-like decrease of $\lambda_{3}$ above $T_{c}$ seen in Fig.~\ref{fig:Ce_PSI_ZFlarget}(b). 

Our measurements provide the following picture of the magnetism in this material.
(i) There is likely a slowing of Ce moments below 30~K, seen via the peak in $\lambda_{3}$, with a possible freezing below $5$~K, which gives rise to an increase in precession frequency;
(ii) The Mn system is likely to be ordered throughout the bulk below 218~K.
In the region $218\leq T\leq 318$~K the magnetic fraction of the material causing resolvable relaxation decreases, with a new relaxation channel causing a fraction of the muon polarization to be rapidly depolarized, such that a component of the asymmetry is not resolved.
This effect possibly reflects the onset of a dynamically fluctuating component to the magnetism since, in the same region, dynamic fluctuations become more prevalent in the longitudinal relaxation channel, where they cause a rapid increase and a peak in $\lambda_{3}$ as the reordering transition at $T_{c}=318$~K is approached from below. 
However, the distribution, $\Delta$, of magnetic fields at the muon sites remains correlated enough over time and length scales such that oscillations can continue to be resolved. 
(iii) Above the spin reordering transition (in the region $318\leq T\leq417$~K), the AFM phase is significantly less well correlated.  The dynamics are rapid on the muon timescale, preventing oscillations from being resolved. Since it is unlikely that the local magnetic field at the muon site
is parallel to the initial spin polarization in this phase, we attribute the lack of oscillations to a qualitative difference in magnetic behaviour between the phases above and below $T=318$~K, with higher-temperature AFM state characterized by rapid fluctuations, and a broader distribution of local fields at the muon sites. This might reflect the presence of ordered domains separated by disordered walls, for example, or intrinsic disorder throughout the bulk of the material.

\section{\pmg}
\label{sec:Pr}
\pmg~is reported to undergo a complicated cascade of magnetic phase transitions~\cite{welter1995neutrons,Kervan2004_Prmagchara,Wang2014PrNeutrons,song2022Pranisotropy}, comprising the following:
\begin{enumerate}
\itemsep-0.25em 
\item[(a)] F$_\mathrm{Pr}$ phase ($T \lesssim$ 80~K), where the Mn spins display an incommensurate conical magnet structure with its axis along $c$ and the Pr spins FM order along $c$.
\item[(b)] I$_{c}$ phase (80 $\lesssim T \lesssim$ 280~K), where the Mn spins form the same incommensurate conical magnet structure along the $c$ axis, but the Pr spins disorder.
\item[(c)] C phase (280 $\lesssim T \leq$ 331~K), where the Mn spins assume a canted AFM structure, with a net ferromagnetic moment along the $c$ axis. 
\item[(d)] AFM phase (331 $\leq T \leq$ 415~K), where the Mn spins display a collinear AFM structure in the $a$-$b$ plane.
\item[(e)] PM phase ($T >$ 415~K), where the Mn spins disorder.
\end{enumerate}
The transitions between these phases are denoted $T_\mathrm{Pr}\approx80$~K, $T_{\mathrm{icm}}\approx280$~K, $T_c=331$~K and $T_\mathrm{N}=415$~K.

\subsection{\musr~results}
\begin{figure*}
  \centering
  \includegraphics[width=0.9\linewidth]{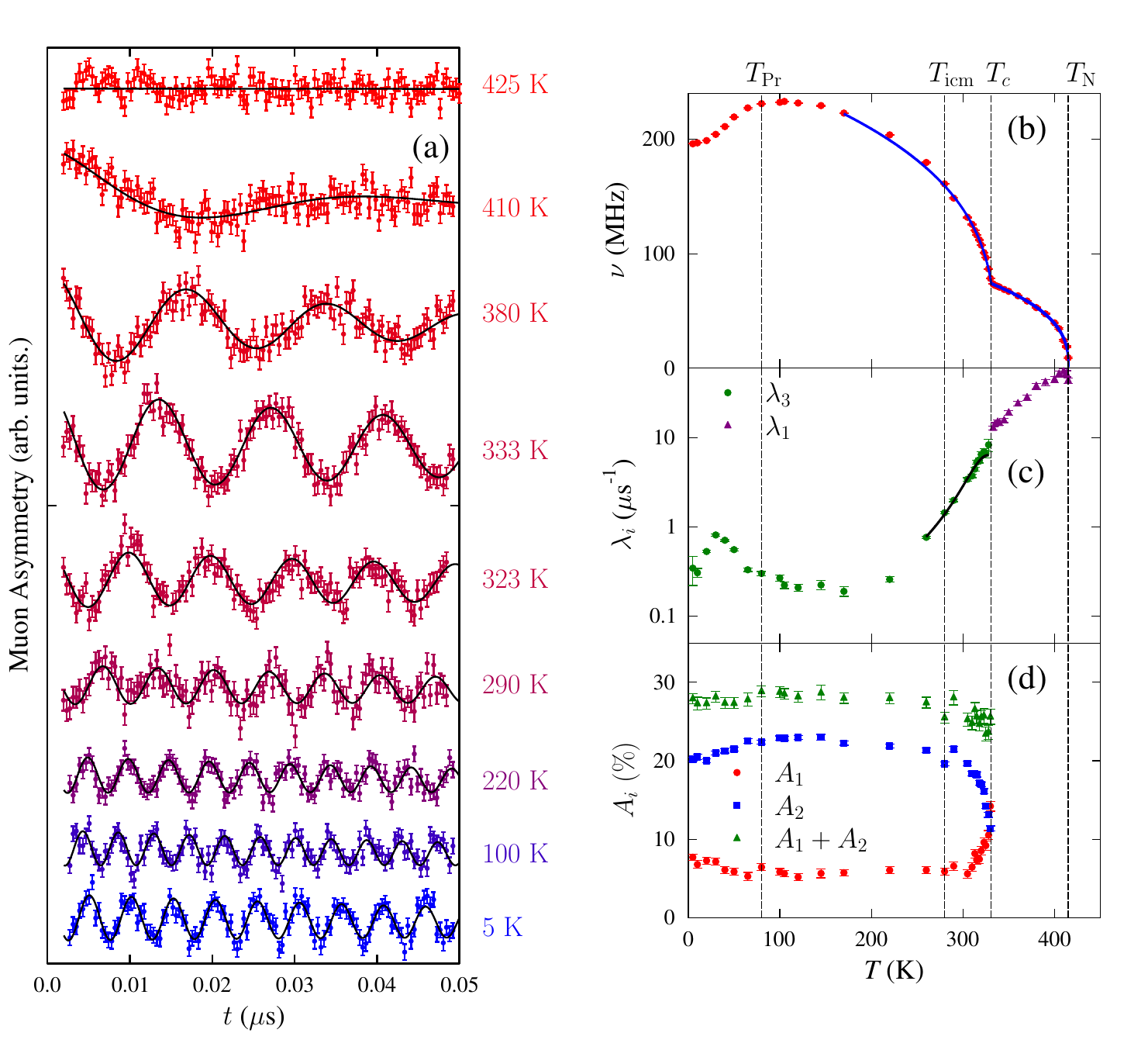}
  \caption{ZF~\musr~measurements of~\pmg. (a) A set of asymmetry spectra, demonstrating the evolution of the oscillations with temperature. (b) Variation of the precession frequency with temperature, fitted with Eq.~\ref{eq:Pr_osc_fit}. (c) Temperature dependence of the slow magnetic dynamics ($\lambda_3$) below $T_c$ (with the fit of Eq.~\ref{eq:Lorentzian_slowrelaxation} shown in black) and the rapid relaxation of the oscillating term ($\lambda_1$) above $T_c$. (d) Variation of oscillating ($A_1$), relaxing ($A_2$) and total amplitude below $T_c$, demonstrating the swapping of amplitudes between $T_{\mathrm{icm}}$ and $T_c$.} 
  \label{fig:Pr_PSI_ZF}
\end{figure*}
All regions reported to host magnetic ordering in~\pmg~give rise to a single oscillating frequency in the ZF~\musr~spectra. The first $0.05~\mu$s of these spectra are captured by the function
\begin{equation}
	A\left(t\right) = A_{1}e^{-\lambda_1t}\cos(2\pi\nu t) + A_{2}e^{-\lambda_2t}. 
    \label{eq:Pr_osc}
\end{equation}
These terms correspond, respectively, to the component of the muon spin initially perpendicular to the local magnetic field, which undergoes rapid precession at frequency $\nu=\frac{\gamma_\mu}{2\pi}B$ and the component of the muon spin initially parallel to the local field, which are slowly relaxed by magnetic dynamics at a rate $\lambda_2$. 
Relaxation rates $\lambda_1=9.1(9)~\rm\mu s^{-1}$ and $\lambda_2=1.9(2)~\rm\mu s^{-1}$ are globally refined and fixed below $T_c=331$~K.
Rather than observing an increase in local field with decreasing $T$, the precession frequency $\nu$ is seen to first plateau below $170$~K, before decreasing below $80$~K [Fig.~\ref{fig:Pr_PSI_ZF}(b)].
This indicates a low-$T$ transition in~\pmg, which has previously been attributed to the magnetic ordering of the Pr ions~\cite{Kervan2004_Prmagchara,Wang2014PrNeutrons}.
Therefore, our~\musr~results agree with reports of Pr ordering up to $150$~K~\cite{welter1995neutrons,huang2023Pranisotropy}.

A transition to a canted ferromagnetic ordering of the Mn ions is reported at $T_{\mathrm{icm}}=280$~K, but we observe no corresponding change in behaviour of the precession frequency $\nu(T)$.
Instead, $T_{\mathrm{icm}}$ marks the onset of the decrease in relaxing amplitude and an increase in oscillating amplitude [Fig.~\ref{fig:Pr_PSI_ZF}(d)], rather like the behaviour seen in the $R$=Ce material. Here the change in amplitudes is likely independent of the loss of incommensurate order, since the similar amplitude changes in~\cmg~can be explained by the reorientation of the Mn spins into the $a$-$b$ plane. 

For $T_c<T<T_\mathrm{N}$, where no oscillations were observed in~\cmg, Eq.~\ref{eq:Pr_osc} still captures the behaviour of the spectra, but now the amplitudes, $A_{1}=16.7(3)~\%$ and $A_{2}=11.1(2)~\%$, and the relaxation rate, $\lambda_2=4.5(6)~\rm\mu s^{-1}$, are globally refined to different values. 
The amplitudes are similar to the fitted values just below $T_c$ [$A_1=14.0(6)$~\% and $A_2=11.2(3)$~\%], suggesting that the spin reorientation occurs gradually for $T<T_c$ rather than suddenly at $T=T_c$. 
In the region where conventional order-parameter behaviour is observed the frequency is fitted with
\begin{equation}
\nu(T) =\begin{cases}
			\nu_1\left(1-T/T_{\tilde{c} }\right)^{\beta_1}, & \text{for~} 170~\mathrm{K}<T<T_c\\
            \nu_2\left(1-T/T_\mathrm{N}\right)^{\beta_2}, & \text{for~}
            T_c<T<T_\mathrm{N}
		 \end{cases}~,
         \label{eq:Pr_osc_fit}
\end{equation} 
where $\nu_2=\nu_1\frac{\left(1-T_c/T_{\tilde{c}}\right)^{\beta_1}}{\left(1-T_c/T_\mathrm{N}\right)^{\beta_2}}$. (We use this approach so that continuous behaviour of the frequency at $T_c$ is captured by the fitting.)
The low temperature region of the fit is parametrised by $\nu_1=275.5(6)$ and $\beta_1=0.30(2)$, with $T_{\tilde{c}}=335.4(1)$~K giving the temperature at which $\nu$ would equal zero if a change in magnetic order did not occur at $T_c=331.2(2)$~K.
Above the change in behaviour at $T_c$, the fit is parametrised by $T_\mathrm{N}=415.3(1)$~K and $\beta_2=0.37(2)$.
The value of $\beta_2$ is similar to that of the of the 3D Heisenberg model ($\beta=0.367$)~\cite{blundell2003magnetism}, suggesting the fluctuations of the Mn spins in the vicinity of the phase transition are Heisenberg-like. 
In the AFM phase ($T_{c}<T<T_{\mathrm{N}}$) an increase in the relaxation of the oscillations (relaxation rate $\lambda_1$) occurs [Fig.\ref{fig:Pr_PSI_ZF}(c)], with a peak value of $54~\mu$s$^{-1}$ at $410$~K ($5$~K below $T_\mathrm{N}$).

In order to examine the magnetic dynamics of~\pmg, the asymmetry below $T_c=331$~K is captured using
\begin{equation}
	A\left(t\right) = A_{3}e^{-\lambda_3t} + A_{4}
    \label{eq:Pr_rel}
\end{equation}
over an $8~\mu$s time window, where a large binning factor is used to average out the rapid early time oscillations.
The relaxation rate, $\lambda_3$, behaves in a similar fashion to that observed for~\cmg, with a peak at low temperatures ($\approx 50$~K), before a roughly exponential increase as the transition to AFM order in the $a$-$b$ plane above $T_{c}$ is approached [Fig.~\ref{fig:Pr_PSI_ZF}(c)]. 
Fitting the peaked behaviour using Eq.~\ref{eq:Lorentzian_slowrelaxation} and a value of $T_\mathrm{peak}=328$~K results in a characteristic width of $\Gamma=25(1)$~K, giving a characteristic energy of $2.2$~meV, similar to the $R$=Ce material.

Above $T_c=331$~K the slower oscillations were fitted using Eq.~\ref{eq:Pr_osc}, as in the shorter time window. 
Throughout the AFM phase there appears to be no significant change in the slow magnetic dynamics, with $\lambda_2$ remaining constant. The increase in the damping rate of the oscillating term $\lambda_{1}$ as $T_\mathrm{N}=415$~K is approached can be attributed to an increase in the width of the distribution of disordered fluctuating fields, similar to that seen in~\cmg. However in the case of~\pmg~the amount of disorder is less, since coherent oscillations are still seen above $T_c$.

\subsection{Muon site analysis}

In the I$_{c}$ phase at $200$~K (which is significantly above any reported Pr-moment ordering), the magnetic structure is characterised by neutron scattering measurements~\cite{Wang2014PrNeutrons,welter1995neutrons}, giving $m_\mathrm{Mn}=2.90(4)\mu_\mathrm{B}$, $\alpha=53.6(8)^\circ$ and $q_z=0.214(1)$, with the helical component of the Mn spins AFM aligned within the $a$-$b$ plane. 
The muon site has a field distribution with an average field of $1450$~mT ($197$~MHz), which overlaps the experimental precession frequency of $209.0(4)$~MHz [Fig.~\ref{fig:R_field_dist}(b)]. 

As the temperature decreases, the Mn moment size and propagation vector increase [$m_\mathrm{Mn}=3.23(7)\mu_\mathrm{B}$ and $q_z=0.272(2)$ at $10$~K], and the moments align closer to the $c$ axis [$\alpha=51.0(9)^\circ$]. 
This is accompanied by FM aligning of the Pr spins along $c$ with moment $m_\mathrm{Pr}=1.53(3)\mu_\mathrm{B}$~\cite{welter1995neutrons,Wang2014PrNeutrons}.
Below $T_{\mathrm{Pr}}$ we observed a decrease in the measured muon precession frequency of $12.2$~MHz (corresponding to a decrease in local field at the muon site of $90$~mT). However our calculations based on this scenario predict an increase in average local field at the muon site of $680$~mT ($91$~MHz), 
caused by the increased size of the Mn moments. We therefore conclude that, despite the success of our approach for the state realized at $200$~K, we are unable to capture the behaviour of the local field when the Pr moments order below $T_{\mathrm{Pr}}$, with the behaviour predicted by our simulations being more consistent with a continuation of the increase in the precession frequency $\nu$ on cooling that we observe from $T_{c}$ down to $\approx 200$~K [Fig.~\ref{fig:Pr_PSI_ZF}(b)]. 
This discrepancy could be due to the published magnetic structure being incorrect. However, alternative explanations (including one specific to~\pmg) are discussed below.

\subsection{Discussion}
The overall picture of magnetism in this material can be summarised as being similar to the Ce material, but with a few key differences. 
The main features are
(i) Pr order at low temperatures, which is more conclusive and persists to much higher temperatures ($170$~K) than Ce ordering in~\cmg; 
(ii) a decrease in the total amplitude below $T_{c}$, along with the onset of dynamics in the region $T_{\mathrm{icm}}\leq T \leq T_{c}$; 
(iii) oscillations persist in the region $T_{c}<T<T_{\mathrm{N}}$, but the relaxation rate of these increases with increasing temperature. 
Although magnetic disorder increases in this region, it is not large enough in this case to wash out the muon-spin oscillations. 

It is interesting that our dipole field calculations (using moment values from neutron diffraction measurements) fail to capture the experimental decrease in muon precession frequency in~\pmg~at low temperatures. 
One possible reason for this is that the Ruderman-Kittel-Kasuya-Yosida (RKKY) interaction~\cite{blundell2003magnetism} can cause ordered Pr moments to polarize nearby conduction electrons, leading to an increased hyperfine contribution to the local field at the muon site. 
Effects of the RKKY interaction have also been reported in other $RM_2X_2$ intermetallics~\cite{bouaziz2022centro,wood2025magnon}.
Additionally Pr$^{3+}$ is a non-Kramers ion, with crystal field calculations on~\pmg~identifying a non-Kramers doublet $9.7$~meV above a ground state singlet [see the Supplementary Material~\cite{sm_muon}].
Muon induced distortions can modify the local environment of the Pr ion, leading to a splitting of the doublet that is not symmetry protected.
This effect can lead to differences in magnetic response of the muonated material compared to the material without the muon impurity~\cite{Tashma1997_PrKramers,Foronda2015_PrKramers}.  
Finally, similar decreases in precession frequency at low temperatures have been reported in~\musr~measurements on single crystal samples of Dy, Ho and Er~\cite{Schreier2000_metallicREcontact}.
Unlike~\pmg~there is only magnetic order due to rare-earth ions in these materials, so spontaneous bulk magnetisation due to magnetic anisotropy was suggested as the reason for the decrease in local field at the muon site.  

\section{\nmg}
\label{sec:Nd}
Previous magnetic characterisation of~\nmg~has identified the most complex series of magnetic phases in the~\remgall~series~\cite{welter1995neutrons,xu2023Ndtopological,Morellon1997_Nd}, which are:
\begin{enumerate}
\itemsep-0.25em 
\item[(a)] F$_\mathrm{Nd}$ phase ($T \leq$ 21 K), where the Mn spins form an incommensurate conical structure with its axis in the $a$-$b$ plane and the Nd spins FM order along the same axis.
\item[(b)] I$_{ab}$ phase (21 $\leq T \leq$ 215~K), where the Mn spins display an incommensurate conical structure with its axis in the $a$-$b$ plane, but the Nd spins disorder. 
\item[(c)] I$_{c}$ phase (215 $\leq T \lesssim$ 240~K), where the Mn spins form an incommensurate conical  structure along the $c$ axis.
\item[(d)] C phase (240 $\lesssim T \leq$ 335~K), where the Mn spins assume a canted AFM structure, with a net ferromagnetic moment along $c$. 
\item[(e)] AFM phase (335 $\leq T \lesssim$ 425~K),
where the Mn spins display a collinear AFM structure in the $a$-$b$ plane.
\item[(f)] PM phase ($T \gtrsim$ 425~K), where the Mn spins disorder.
\end{enumerate}
The transitions between these phases are labelled $T_\mathrm{Nd}=21$~K, $T_\mathrm{SR}=215$~K, $T_\mathrm{icm}\approx240$~K, $T_\mathrm{c} = 335$~K and $T_\mathrm{N} \approx 425$~K. 
\begin{figure}
  \centering
  \includegraphics[width=0.9\linewidth]{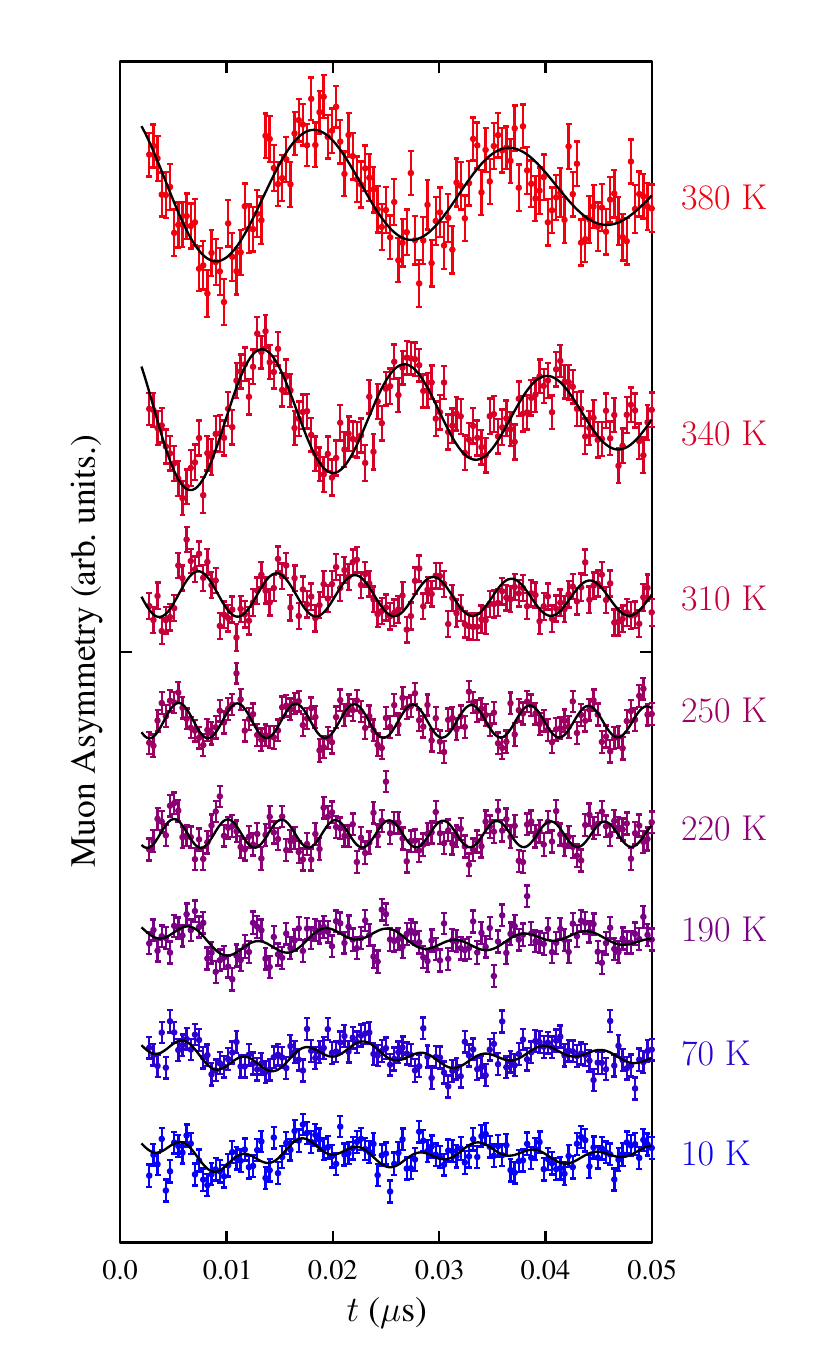}
  \caption{Evolution of the early-time spectra of~\nmg~with temperature, demonstrating the transition to a single frequency above $T_\mathrm{SR}=215$~K that increase in amplitude with spin-reorientation into the $a$-$b$ plane at higher temperatures.}
  \label{fig:Nd_PSI_ZFspectra}
\end{figure}

Along with this cascade of magnetic transitions, a phase diagram of~\nmg~produced using LTEM in Ref.~\cite{hou2021skyrmions} highlights two distinct regions hosting topological excitations.
Between temperatures of $220$~K and $250$~K a region containing a combination of SkBs, bubbles and stripes is seen in fields up to $120$~mT, but with a relatively low density of excitations. In contrast, the same field range between $250$~K and $320$~K hosts a high density region of SkBs. 
Moreover, the skyrmionic core polarity cannot be reversed by the application of a magnetic field~\cite{treves2025Ndstability}, highlighting the absence of the DM interaction and reaffirming the excitations in~\nmg~are skyrmion bubbles.

\subsection{Continuous-source~\musr}
Oscillations are observed across the entire temperature range, matching the reports of magnetic ordering up to $425$~K~\cite{boschsantos2015_Ndhf,nowik1995_Ndmag,Morellon1997_Nd}. 
Clear changes in the form of these oscillations is observed [Fig.~\ref{fig:Nd_PSI_ZFspectra}], corresponding to reported transitions in magnetic structure, as described below. 
\subsubsection{F$_\mathrm{Nd}$ and I$_{ab}$ phases}
At temperatures $T\leq215$~K, the spectra were fitted with the function
\begin{equation}
\label{eq:Nd_two_osc}
	A\left(t\right) = \sum_{i=1}^{2}A_{i}e^{-\lambda_it}\cos\left(2\pi\nu_{i}t+\phi_{i}\right) + A_3,
\end{equation}
where the parameters are globally refined and fixed to $A_1=2.4(2)$~\%, $A_2=2.9(2)$~\%, $\lambda_1=17(2)~\rm\mu s^{-1}$, $\lambda_2=24(3)~\rm\mu s^{-1}$, $\phi_1=-30(4)^{\circ}$ and $\phi_2=-51(4)^{\circ}$. 
The presence of two oscillatory components suggests the existence of  two magnetically distinct muon sites in~\nmg~(we discuss evidence for a third below). 
Multiple magnetically-distinct sites are not resolvable in the other materials in the~\remg~series and the additional precession frequencies become resolvable following the spin reorientation of the incommensurate structure into the $a$-$b$ plane at temperatures below $T_{\mathrm{SR}}$, which is a unique feature of this material. 

\begin{figure}
  \centering
  \includegraphics[width=0.9\linewidth]{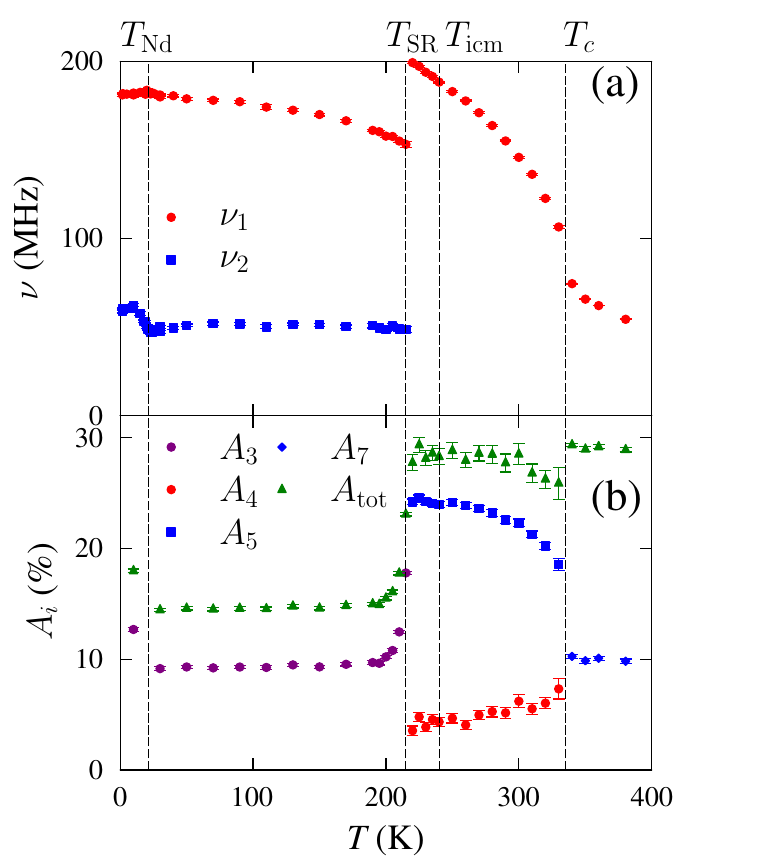}
  \caption{Fits of zero-field muon measurements of~\nmg, demonstrating the change in (a) oscillating frequencies and (b) amplitudes with temperature. A discontinuous change from two to one precession frequencies can be seen at $T_\mathrm{SR}$=215~K, accompanied by an increase in total asymmetry, $A_\mathrm{tot}$.   }
  \label{fig:Nd_PSI_ZF}
\end{figure}

Within this temperature regime we observe a transition attributed to the magnetic ordering of the Nd sublattice at $T_\mathrm{Nd} = 21$~K. This is indicated by an increase in the smaller precession frequency $\nu_2$ [Fig.~\ref{fig:Nd_PSI_ZF}(a)] and the constant contribution to the asymmetry $A_3$ [Fig.~\ref{fig:Nd_PSI_ZF}(b)].
The lack of change in $\nu_1$ suggests the local field at only one of the muon sites 
is significantly altered by the magnetic ordering of the Nd ions [Fig.~\ref{fig:Nd_PSI_ZF}(a)].

\begin{figure}
  \centering
  \includegraphics[width=0.9\linewidth]{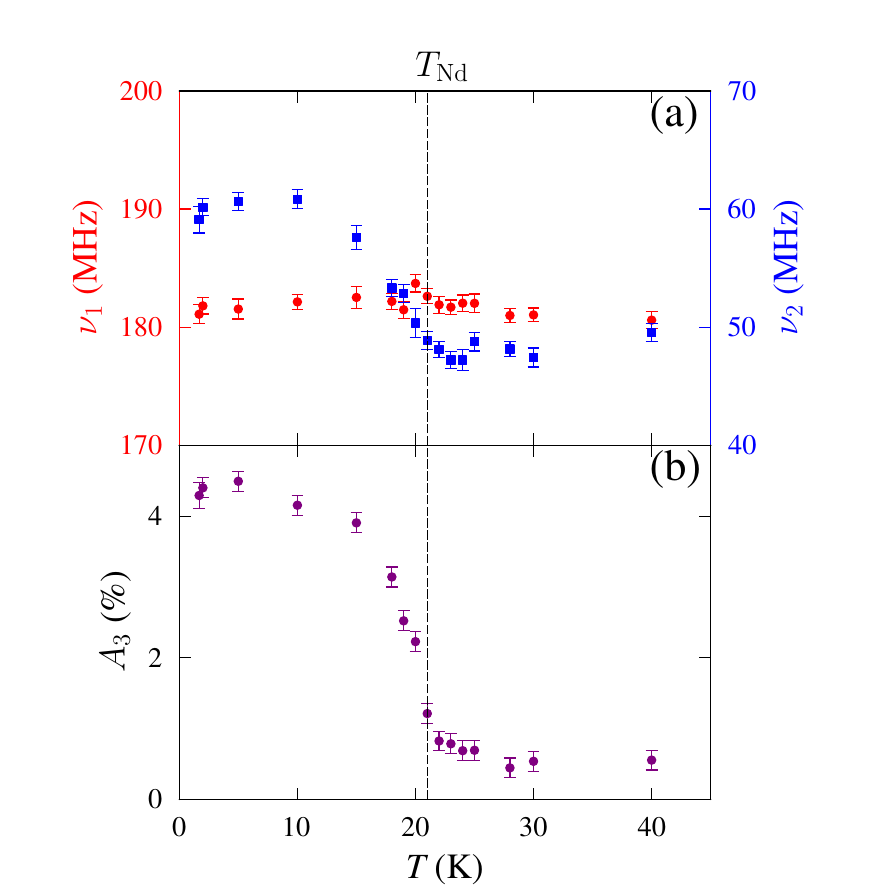}
  \caption{Fits of the low temperature ZF muon measurements on~\nmg, demonstrating the (a) difference in the behaviour of the oscillating frequencies across the Nd ordering transition and (b) the change in the constant contribution to $A(t)$.}
  \label{fig:Nd_lowT_PSI_ZF}
\end{figure}

Inelastic neutron scattering~\cite{chatterji2012_Ndinelastic}~suggests that the ordering transition of the Nd ions occurs over a wide temperature range as is the case in~\pmg, however this is not reflected in our~\musr~and other magnetometry and neutron diffraction measurements~\cite{xu2023Ndtopological,wang2020Ndskyrmions}. 
In fact, our muon measurements at low temperature [Fig.~\ref{fig:Nd_lowT_PSI_ZF}] indicate a continuous variation in $\nu_{2}$ across the transition and, as $\nu_{2}$ can be treated as an order parameter, this suggests the ordering of the Nd sublattice is a continuous transition with a critical temperature of $21$~K. 

The I$_{ab}$ phase is unique to the Nd-containing material, being absent from the Ce- and Pr-containing analogues.
In the $\mathrm{I}_{ab}$ phase above the Nd ordering transition, $\nu_{2}$ remains relatively unchanged, while $\nu_{1}$ gradually decreases as $T_\mathrm{SR}$ is approached. 
The relaxation rates in this phase are found to be temperature independent, suggesting little change in the static and dynamic disorder across the I$_{ab}$ phase. 
Notable, however, in this phase is a large amount of missing asymmetry which is recovered at higher temperatures. 
One possible explanation for this is that the amplitude from a distance class of muon site is relaxed too quickly to be resolved, as discussed below.

\subsubsection{I$_{c}$ and C phases}
Above the spin reorientation transition at $T_{\mathrm{SR}}$ (that is, in the region $215\leq T\leq 335$~K) the spectra were fitted with the function
\begin{equation}
	A\left(t\right) = A_{4}e^{-\lambda_4t}\cos\left(2\pi\nu_1 t\right) + A_{5}e^{-\lambda_2t},
\end{equation}
where $\lambda_4=7(2)~\rm\mu s^{-1}$ and $\lambda_5=1.2(2)~\rm\mu s^{-1}$ are globally refined and fixed.
The single oscillating term suggests that the muon sites become magnetically indistinguishable above $T_\mathrm{SR}$, and the sudden switch from two distinct precession frequencies to a single frequency implies that $T_\mathrm{SR}$ is a discontinuous transition.
Comparing amplitudes in Fig.~\ref{fig:Nd_PSI_ZF}(b), we see a sizeable recovery in the total initial asymmetry, $A(t=0)$, across the spin reorientation transition.
Changes in magnetic dynamics are captured by the slowly relaxing component with amplitude $A_5$ and relaxation rate $\lambda_{5}$, reflecting muon spin components initially aligned parallel to the internal field direction~\cite{muontextbook2022}.
Increasing in temperature, the reported transition from an incommensurate helical spin configuration to a canted AFM along the same axis at $T_\mathrm{icm} \approx 240$~K is not apparent in the precession frequency (as in~\pmg), suggesting the local field at the muon site remains unchanged. 
The behaviour in this phase resembles the features seen in the other two materials in the series below $T_{c}$. As in these previous cases we see a small decrease in the total relaxing asymmetry, with a slight increase in the oscillating amplitude accompanied by a clearer decrease in the relaxing amplitude. 

\subsubsection{AFM phase}
The spectra in the AFM phase ($335\leq T\leq 380$~K) were fitted using
\begin{equation}
	A\left(t\right) = A_6e^{-\lambda_6t}\cos\left(2\pi\nu_{1}t\right)+A_7,
\end{equation}
where $A_6=19.1(7)\%$ and $\lambda_6=19.3(10)~\rm\mu s^{-1}$ are globally refined. 
The transition to collinear AFM order in the $a$-$b$ plane leads a relative increase in the amplitude of the oscillating component ($A_6$ compared to $A_4$) and a reduction in the slowly relaxing and constant contributions to $A(t)$.
Moreover, the precession frequency $\nu_1$ continues to decrease with temperature, but the rate of change sharply decreases at $T_c=335$~K, as it did in the Pr-containing system. 

\subsection{Pulsed-source~\musr}
Measurements on the same mosaic sample were also made at ISIS. As ISIS is a pulsed source this allows us to access a longer time window, so it is ideal for investigating slow magnetic dynamics. The width of the pulse in time means that we are unable to resolve rapid oscillations associated with magnetic ordering (similar to the impact of our heavy binning of the~\cmg~and~\pmg~data sets, discussed above), so the spectra were fitted to 
\begin{equation}
	A\left(t\right) = A_{8}e^{-\lambda t} + A_9,
    \label{eq:Nd_ISIS_relax}
\end{equation}
where $A_8$ corresponds to the muons relaxed in magnetic environments and $A_9$ is a constant contribution corresponding to muons implanting in non-magnetic regions, such as the silver packet. 

\begin{figure}
  \centering
  \includegraphics[width=0.9\linewidth]{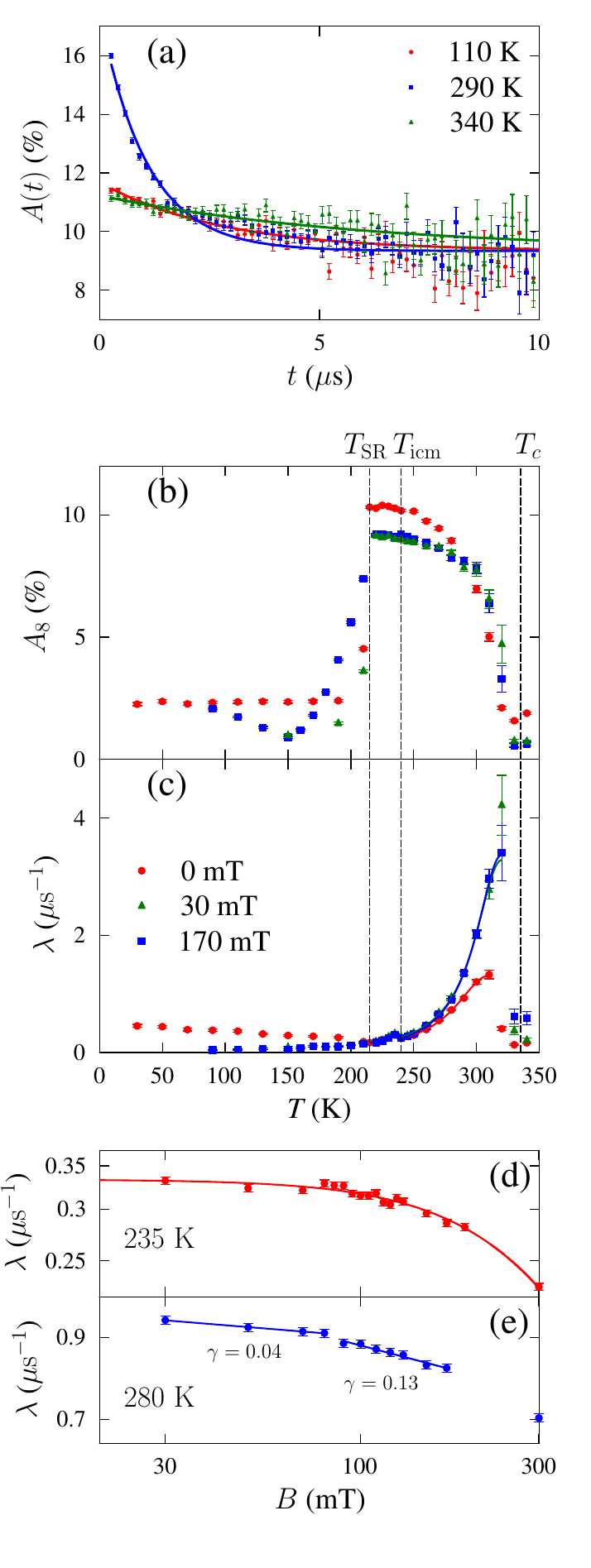}
  \caption{(a) Example spectra of ZF measurements made on~\nmg~at ISIS, where temperature scans show the change in (b) relaxing asymmetry and (c) relaxation rate with temperature at different applied longitudinal magnetic fields. Field scans demonstrate the variation of relaxation rate, $\lambda$ with applied magnetic field at (d) $235$~K and (e) $280$~K, with fits described in the text.}
  \label{fig:Nd_ISIS_LF}
\end{figure}

In measurements made at zero applied field, transitions can be seen at $T_\mathrm{SR} = 215$~K and $T_c = 335$~K in the relaxing asymmetry [Fig.~\ref{fig:Nd_ISIS_LF}(b)]. 
The relaxation rate $\lambda$, which quantifies the magnetic dynamics in~\nmg, slowly increases between $T_\mathrm{SR}=215$~K and $T_\mathrm{icm}=240$~K, before rapidly increasing and reaching a maximum value just below the transition to in-plane antiferromagnetism at~$T_c$. 
This behaviour was fitted using Eq.~\ref{eq:Lorentzian_slowrelaxation} and $T_\mathrm{peak}=310$~K for ZF, resulting in a width of $\Gamma=33(1)$~K which corresponds to a characteristic energy of $2.8$~meV.
The transition at $T_\mathrm{icm}\approx240$~K is seen via a small peak in $\lambda$, suggesting that the loss of incommensurate magnetic order in~\nmg~impacts the slow magnetic dynamics.

The continuous and pulsed source~\musr~measurements together give a picture of the material that can be compared with our observations from the $R$=Ce and Pr systems:
(i) Nd ordering is observed at low temperature; 
(ii) a discontinuous change in magnetic structure at $T_{\mathrm{SR}}$; 
(iii) a decrease in amplitude below $T_{c}$, along with the onset of dynamics in the region $T_{\mathrm{SR}}\leq T \leq T_{c}$; 
(iv) A feature in the magnetic dynamics corresponds to the loss of incommensurate magnetic order at $T_\mathrm{icm}$;  
(v) Oscillations persist in the region above $T_{c}$.
 
\subsection{LF measurements}

In order to probe the effect of applied fields on this material, we made a series of longitudinal-field (LF) measurements at both continuous and pulsed sources.  Measurements in a LF of $B_0=170$~mT were made at the continuous source at PSI. The primary impact of this applied field is to broaden the magnetic transitions, which is particularly apparent at $T_\mathrm{SR}=215$~K. 
Similar measurements were made at the pulsed source at ISIS in LFs of $B_0=30$~mT and $170$~mT, where the applied field also broadens the transition region around $T_\mathrm{SR}=215$~K, as seen in the amplitudes [Fig.~\ref{fig:Nd_ISIS_LF}(b)].   
In both cases the broadening can be attributed to the applied field along the $c$ axis making the spin reorientation more energetically favourable, so it commences at lower temperatures.

In the $T$ regime just below $T_{c}$,
the magnitude of the relaxation is enhanced by field, with the greatest relaxation in the SkB-hosting region of the phase diagram (at $30$~mT).

The peak in relaxation slightly increases in temperature with applied field, which can be prescribed to the greater energy cost of rotating spins into the $a$-$b$ plane with an applied field along the $c$ axis. 
Fitting to the previous model [Eq.~\ref{eq:Lorentzian_slowrelaxation}], we find $T_\mathrm{peak}=320$~K for $B_0=30,~170$~mT with a width of $\Gamma=26(1),~25(1)$~K ($\approx2.2$~meV) for $B_0=30,~170$~mT.

The magnetic dynamics of~\nmg~were then further investigated at ISIS by performing a pair of field scans (at $235$~K and $280$~K) to probe different parts of the SkB lattice.
The resulting spectra were observed to decay with exponential relaxation, so were fitted using Eq.~\ref{eq:Nd_ISIS_relax} with both $A_8$ and $A_9$ globally refined and fixed.
For dynamic fluctuations of a Gaussian distribution of magnetic field components the longitudinal relaxation rate, $\lambda$, is described by the Redfield equation~\cite{muontextbook2022,Hernandezmelian2023_organicLF}, 
\begin{equation}
	\lambda = \frac{2\Delta^2\tilde\nu}{\tilde\nu^2+\gamma_{\mu}^2B_{0}^2},
\end{equation}
where $B_0$ is the magnitude of the applied longitudinal magnetic field, $\tilde\nu$ and $\Delta$ are the rate and amplitude of fluctuations respectively and $\gamma_{\mu} = 2\pi\times135.53$ MHz T$^{-1}$ is the muon gyromagnetic ratio.
Fitting to the model at $235$~K, we find $\tilde\nu=373(8)$~MHz and $\Delta=7.89(7)$~MHz, and at $280$~K, $\tilde\nu=438(12)$~MHz and $\Delta= 14.2(2)$~MHz, confirming we are in the fast fluctuation limit.  
However, significantly better agreement to the Redfield equation is found at $235$~K [see~\ref{fig:Nd_ISIS_LF}(d)].
Recalling the phase diagram of~\nmg~\cite{hou2021skyrmions}, the field scan at $235$~K passes through multiple phases hosting a range of magnetic objects while the scan at $280$~K only passes through a SkB-hosting region, with a significantly greater density of objects at $280$~K. 
Deviations from the Redfield model can occur due to the distribution of magnetic field components no longer being Gaussian, which could be driven by the greater density of SkBs at $280$~K. 
Therefore, considering a parametrization of the form $\lambda\propto B^{-\gamma}$, and plotting $\ln(\lambda)$ against $\ln(B)$ highlights two regions of differing behaviour at $280$~K [Fig.~\ref{fig:Nd_ISIS_LF}(e)]. 
At low fields, $B<90$~mT, we find that $\gamma=0.04(2)$ (i.e.\ a very weak field dependence), while at higher fields, $90\leq B\leq 170$~mT we find $\gamma=0.13(1)$. 
This switch in behaviour corresponds to a large decrease in the reported density of SkBs~\cite{hou2021skyrmions},

\subsection{Muon site analysis}

In the I$_{ab}$ phase the magnetic structure revealed at $150$~K by neutron scattering~\cite{welter1995neutrons} is characterised by $m_\mathrm{Mn}=2.7\mu_\mathrm{B}$, $\alpha=58^\circ$, and  
$q_z=0.183(2)$, with the axis of the conical arrangement of Mn moments orientated within the $a$-$b$ plane.
The local field distribution computed for the muon site [Fig.~\ref{fig:R_field_dist}(d)] has two peaks in field, at $370$ and $1230$~mT ($50$ and $167$~MHz), matching the measured two-frequency response well. 
The switch to a single-frequency~\musr~response above $T_\mathrm{SR}=215$~K is also seen in the local field distribution [Fig.~\ref{fig:R_field_dist}(d)], and the component of this distribution along $c$ is negligible in the I$_{ab}$ phase, explaining the loss of the slow relaxing term in the measurement. 

Below the Nd-ordering temperature at $T_\mathrm{Nd}=21$~K, the magnetic structure at $2$~K is characterised by $m_\mathrm{Mn}=2.7\mu_\mathrm{B}$, $\alpha=56^\circ$, 
$q_z=0.227(1)$ and $m_\mathrm{Nd}=2.35(24)\mu_\mathrm{B}$, with the Nd moments ordered ferromagnetically along the conical axis of the Mn structure~\cite{welter1995neutrons}.
(This is similar to the low temperature change in magnetic order reported in~\pmg, but with no increase in $m_\mathrm{Mn}$.)
A key feature of this transition in the measurement is that it leads to an increase in the smaller precession frequency equivalent to a local field of $60$~mT.
Considering the field distribution at the muon site for a conical axis directed along the $a$ and $b$ directions, the lower-temperature peak in the field distribution is seen to decrease by $50$~mT for an axis along $a$, but increases by $90$~mT for an axis along $b$.
Therefore, magnetic order with an axis along the $b$ direction below $T_\mathrm{SR}$ agrees better with the observed increase of the smaller precession frequency.

In the AFM phase above  $T_c=335$~K the local field at the muon site is orientated almost entirely within the $a$-$b$ plane, consistent with the absence of a slowly relaxing term in the measurement.

\subsection{Discussion}
The magnetism in~\nmg~is similar to the Ce- and Pr-based materials above $T_\mathrm{SR}=215$~K, where the main features are 
(i) a small peak in the dynamic magnetic relaxation rate, coincident in this case with $T_\mathrm{icm}=240$~K; 
(ii) a decrease in amplitude below $T_{c}=331$~K, along with the onset of dynamics in the region $T_{\mathrm{icm}}\leq T \leq T_{c}$; 
(iii) oscillations that persist in the region $T_{c}<T\leq380$~K, but with no slowly relaxing term in the asymmetry due to the local field direction. 
Moreover, the relaxation rate of the oscillating term is approximately constant, suggesting that the magnetic disorder due to dynamic fluctuations is less prevalent in the~\nmg~material.

In contrast, below $T_\mathrm{SR}$, the magnetism is significantly different to the other~\remg~materials, and is characterised by
(i) the spin reorientation of magnetic structures from along the $c$ axis to along the $b$ axis, leading to the observation of two muon precession frequencies (matching the calculated field distributions for the muon site);
(ii) Nd order at low temperatures, $T_\mathrm{Nd}=21$~K, causing an increase in the smaller precession frequency.

The loss of asymmetry below $T_\mathrm{SR}=215$~K could be due to the realisation of an additional magnetically distinct class of muon site that is relaxed too quickly to be resolved, but the magnetic field distribution calculated at our candidate muon site suggests such a site is not intrinsic to the $\mathrm{I}_{ab}$ phase. 
Moreover, both of the oscillating components below $T_\mathrm{SR}$ have smaller amplitudes than would be expected for a local field distribution orientated within the $a$-$b$ plane.
The experimental results could be explained by the realisation of two different types of magnetic behaviour in spatially distinct regions.
The first type of regions are magnetically ordered as a conical helix of Mn moments along the $b$ axis, resulting in the two-frequency response seen in our~\musr~measurements and dipole field calculations. 
In contrast, the second type of regions host magnetic fluctuations that rapidly depolarize any muons that implant in it, leading to a fraction of asymmetry that is relaxed too quickly to be resolved.
Muons implanting in these fluctuating regions experience different magnetic environments to muons implanting in the regions of conical order, which may be due to the formation of magnetic defects (such as domain walls).
Based on our~\musr~measurements, we suggest these fluctuating regions only form when the conical helix of Mn moments reorientates along the $b$ axis.
This explains the absence of similar regions in the $R=$~Ce and Pr materials, and the slower loss of asymmetry when a LF is applied along the $c$ axis (making the spin reorientation less energetically favourable).
Finally, the ordering of the Nd moments below $T_\mathrm{Nd}=21$~K recovers $\approx3.5$~\% of asymmetry [Fig.~\ref{fig:Nd_lowT_PSI_ZF}(b)], suggesting the magnetic fluctuations that rapidly depolarize the muons are enhanced by disordered Nd moments. 
 
\section{Discussion}
\label{sec:Discussion}
To provide an overview of the magnetism of~\remgall~we can now compare and contrast the magnetic behaviour of the materials as temperature is varied.

At low temperatures, ordering of the rare-earth ions is seen in each system. 
This is subtle for the Ce-based material, but more clear for the others. 
In terms of the rare-earth magnetism the Pr-based system is unique, since it leads to a reduction in local field at the muon site that cannot be captured by our dipole field calculations.

At temperatures above the regions of rare-earth ordering a different number of magnetically ordered phases is seen in each material, but the overall trend is from incommensurate canted antiferromagnetism, to commensurate canted antiferromagnetism, to a collinear antiferromagnet as temperature is increased. 
The intermediate transition to commensurate canted antiferromagnetism is not reported in the Ce material, but the change in behaviour around $210$~K closely resembles the change seen at $T_{\mathrm{icm}}$ in the other two systems. 
This series of magnetic transitions is accompanied by an onset of dynamic fluctuations and an increase in magnetic disorder, which are defining features of the magnetism in this series.
Moreover, we might expect the canted magnetic order to be compatible with the twisting nature of SkB-type objects, that have been reported in this temperature region~\cite{hou2021skyrmions}. 
There is a peak in the dynamic relaxation just below the transition to collinear antiferromagnetism in each of these materials, with a characteristic width of $\approx25$~K ($2.2$~meV). 
However, it is hard to give a specific interpretation to this energy scale.

In the~\nmg~material, an applied magnetic field leads to the broadening of the magnetic transitions at $T_\mathrm{SR}$ and $T_c$, and pushes the peak in dynamic relaxation rate to higher energies. 
At temperatures well below this peak in the Nd material, the dynamics are characteristic of random fluctuations. 
However, in the region of the peak the dynamic fluctuations are no longer random. 
This suggests a higher degree of correlation in the regions where a high density of SkB excitations is observed. 

In the $R=$Ce and Nd materials, changes in the~\musr~response due to rare-earth ordering are explained by the local field distribution at the muon site, which also capture the change from a two-frequency to a single-frequency response at $T_\mathrm{SR}$ in~\nmg.
While two peaks are seen in the field distribution of I$_c$ phase in the $R=$Ce and Pr, these peaks are close enough in field that modest broadening due to magnetic fluctuations would be expected to result in the observed single-frequency response.

\section{Conclusion}
\label{sec:Conclusion}
In conclusion,~\musr~measurements on the~\remgall~rare-earth based intermetallics allow us to identify multiple changes in local magnetism with temperature. 
A defining feature of the magnetism in these materials is the switching on of magnetic dynamics as the magnetic order goes from being canted along the $c$ axis to AFM ordered in the $a$-$b$ plane.
This allows for the realisation of magnetic excitations in the form of skyrmion bubbles, which lead to correlated magnetic fluctuations in~\nmg.
The switch on of magnetic dynamics is accompanied by an increase in magnetic disorder which continues into the AFM phase: for~\cmg~this is severe enough that oscillations are no longer resolved.
At low temperatures we see evidence of rare-earth ordering in all the materials, which agree with changes in field distribution at the identified muon stopping site (apart from in~\pmg).
Moreover, the switch to a two-frequency~\musr~response unique to~\nmg~is also seen in the local field distribution at this site.

Our work highlights the wide range of complex magnetic states that can be realised in the rare-earth based intermetallics, and ability of the~\musr~technique to capture changes in long-range magnetic order and the impact of unusual excitations on magnetic dynamics.\\

\section{Acknowledgements}
Part of this work was carried out at the ISIS Neutron and Muon source, Rutherford Laboratory, UK, and S$\mu$S, Paul Scherrer Institute, Switzerland and we are grateful for the provision of beamtime. We also acknowledge travel support and (for A.H.-M) studentship support  from STFC-ISIS. 
This work is supported by EPSRC (UK) under grant EP/Z534067/1 and
also via studentship support for T.~L.~B and A.~H.-M.
N.~P.~B acknowledges the support of the Durham Doctoral Scholarship.
B.~M.~H is supported by UK Research and Innovation (UKRI) under the UK Government’s Horizon Europe funding guarantee [Grant No. EP/X025861/1].
Research data will be made available via \textcolor{red}{XXX}. 

\bibliography{bib}

\end{document}

\title{Supplemental material for ``Local magnetic properties of the rare-earth intermetallics~\remgall''}

\author{N.~P.~Bentley}
\affiliation{Department of Physics, Centre for Materials Physics, Durham University, Durham, DH1 3LE, United Kingdom}

\author{T.~L.~Breeze}
\affiliation{Department of Physics, Centre for Materials Physics, Durham University, Durham, DH1 3LE, United Kingdom}

\author{A.~Hern{\'a}ndez-Meli{\'a}n}
\affiliation{Department of Physics, Centre for Materials Physics, Durham University, Durham, DH1 3LE, United Kingdom}

\author{M.~J.~Pearce}
\affiliation{Department of Physics, Centre for Materials Physics, Durham University, Durham, DH1 3LE, United Kingdom}

\author{T.~J.~Hicken}
\affiliation{PSI Center for Neutron and Muon Sciences, 5232 Villigen PSI, Switzerland}

\author{M.~T.~F.~Telling} 
\affiliation{ISIS Neutron and Muon Source, STFC Rutherford Appleton Laboratory, Harwell, Didcot OX11 0QX, United Kingdom}

\author{B.~M.~Huddart}
%\affiliation{Department of Physics, Centre for Materials Physics, Durham University, Durham, DH1 3LE, United Kingdom}
\affiliation{Clarendon Laboratory, University of Oxford, Department of Physics, Oxford OX1 3PU, United Kingdom}

\author{G.~D.~A.~Wood}
\altaffiliation[Current address: ]{ISIS Neutron and Muon Source, STFC Rutherford Appleton Laboratory, Harwell, Didcot OX11 0QX, United Kingdom}
\affiliation{Department of Physics, University of Warwick, Coventry, CV4 7AL, United Kingdom}

\author{D.~A.~Mayoh}
\affiliation{Department of Physics, University of Warwick, Coventry, CV4 7AL, United Kingdom}

\author{G.~Balakrishnan}
\affiliation{Department of Physics, University of Warwick, Coventry, CV4 7AL, United Kingdom}

\author{S.~J.~Clark}
\affiliation{Department of Physics, Centre for Materials Physics, Durham University, Durham, DH1 3LE, United Kingdom}

\author{T.~Lancaster}
\affiliation{Department of Physics, Centre for Materials Physics, Durham University, Durham, DH1 3LE, United Kingdom}

\begin{abstract}
  In this Supplemental Material we present additional details of the muon site analysis performed on~\remgall. We discuss the candidate muon sites identified in each of these materials, and calculate the local field at these sites for the reported magnetic phases. In each case this discussion justifies the selection of site A as the muon site likely realised in the~\remg~series of materials.
  We also present calculated crystal field splitting for the rare-earth ion energy levels in each of the materials, highlighting the presence of a non-Kramers doublet in~\pmg. 
\end{abstract}

\maketitle

\section{Muon site calculations}

\subsection{\cmg}
\begin{table}
\begin{tabular}{c|c|c|c}
       Site & Coordinates & Relative energy (eV) & Wyckoff Site ~~\\
       \hline
         A & (0, 0, 0.2) & 0.0 &$4e$~\\
         B & (0.17, 0.15, 0.18) & 0.11 &$32o$~\\
         C & (1/2, 0, 0.1) & 0.17 &$8g$~\\
         D & (0.46, 0.39, 0) & 0.82 &$16l$~\\
\end{tabular}
\caption{Location, relative energies and symmetry of candidate muon sites calculated for~\cmg.}
\label{tab:sym_muonsite_Ce}
\end{table}
Our muon site calculations identify four classes of muon sites in~\cmg~that we call A, B, C and D [see Tab~\ref{tab:sym_muonsite_Ce}]. 
The relative energies are given by the average energy of each class of muon site compared to that of the lowest-energy class, with similar analysis reported for~\pmg~and~\nmg.
Sites A and B sit in close proximity to four and two Mn ions respectively, so experience the largest local fields of the candidate muon sites. 
Site C is around $1.7$~\AA~from a single Mn ion. Finally, site D sits between two Ge ions, greater than $3$~\AA~from the nearest Mn ion.

\begin{figure}
  \centering
  \includegraphics[width=0.9\linewidth]{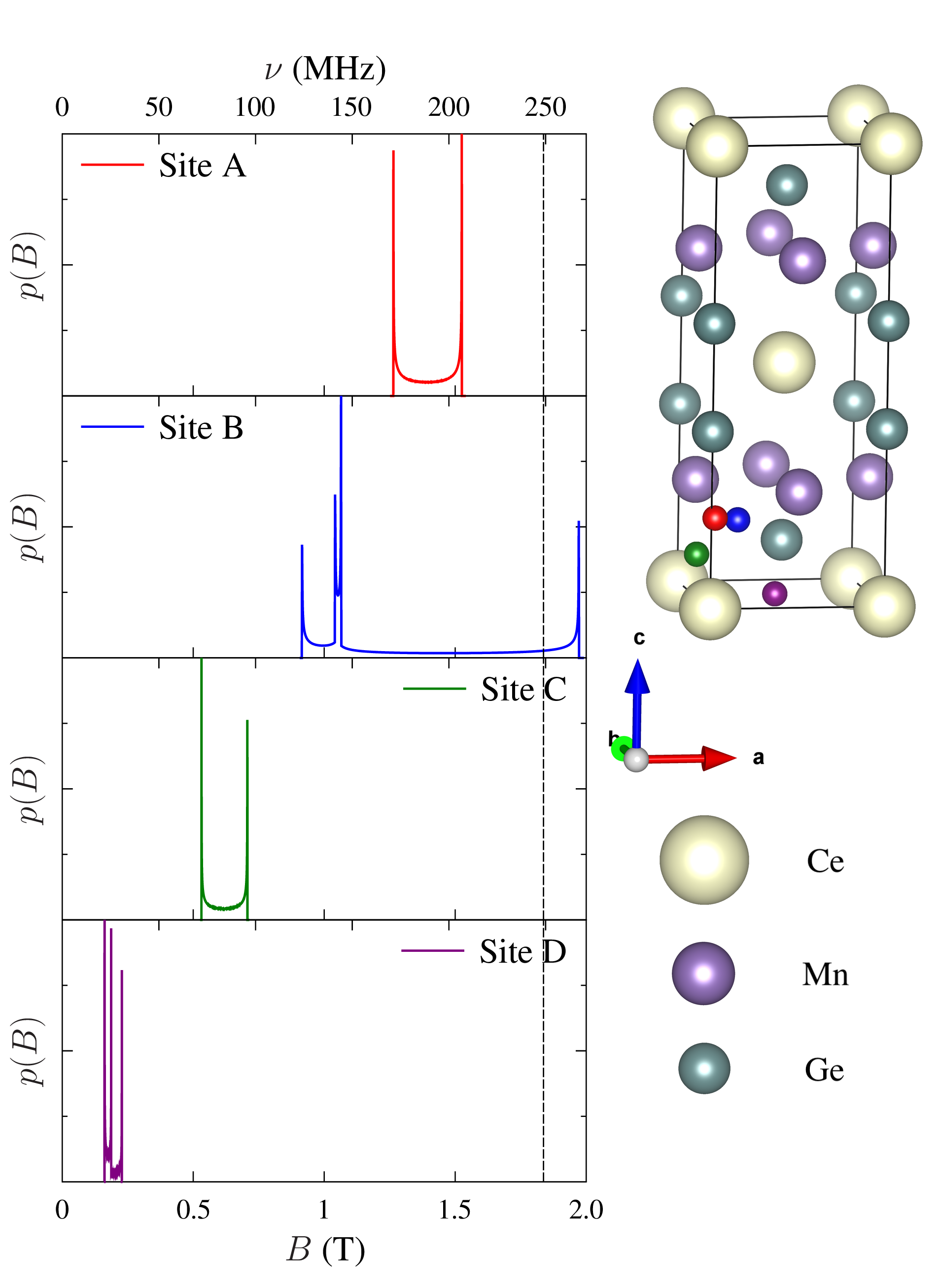}
  \caption{Dipolar field distributions calculated at each of the candidate muon sites in~\cmg~for the measured helical magnetic structure at $4$~K~\cite{MdDin2015Cetunable}, along with a visualisation of the locations of the sites in the unit cell. The dotted line indicated the measured precession frequency, $249.0(5)$~MHz, in our~\musr~measurements. }
  \label{fig:Ce_field_dist}
\end{figure}

The conical magnetic structure reported at $4$~K from neutron scattering measurements on~\cmg~is characterised by $m_\mathrm{Mn}=3.17\mu_\mathrm{B}$, $\alpha=53^\circ$ and $q_z=0.317$, with the helical component of the Mn spins AFM aligned within the $a$-$b$ plane~\cite{welter1995neutrons,MdDin2015Cetunable}. The local field at the muon sites is shown in Fig.~\ref{fig:Ce_field_dist}. 
For~\cmg~we note that, compared to the measured field, the computed field is underestimated in our simulation at all of the sites, with a particularly large discrepancy for sites C and D. 
Therefore, we can suggest it is probable that sites C and D are not realised in~\cmg, which for site D is also supported by the large relative energy of the muon site.

The experimental observation that we resolve only a single-frequency in the~\musr~frequency distributions suggests the dipolar field distributions are broadening into a single broad peak in reality.
This is a more reasonable suggestion for site A which, with a distribution width of $260$~mT, is more than four times narrower than the field distribution for site B.
Moreover, when considering the impact of Ce ion ordering on the local field distribution at the muon site, we find that the field distribution at site B remains unchanged with a small Ce ordered moment ($m_\mathrm{Ce}<1\mu_\mathrm{B}$), in contrast to the low temperature behaviour in the~\musr~measurements.

In the AFM phase above $318$~K, where the Mn spins alternate orientation within the $a$-$b$ plane and along the $c$ axis, the Mn moment size is reported as $\approx2.0\mu_\mathrm{B}$ at $350$~K~\cite{MdDin2015Cetunable}.  
As the direction of the Mn spins in the $a$-$b$ plane is unknown, we calculated the local field for a number of possible orientations.
For sites A, B and C a significant component of the local field is within the $a$-$b$ plane for all the orientations [see Tab.~\ref{tab:Ce_AFM_fields} for moments along the $b$ direction]. This means that absence of oscillations in this phase is not due to the orientation of the local field along the initial muon spin direction, leading to the observation of only the dynamical relaxation of the muon spin (as is the case for the zig-zag phase of Fe$_{1/3}$NbS$_{2}$~\cite{bentley2025intercalated}).
In the conical magnetic state below $318$~K, the component of the field distribution along the $c$ axis is larger for sites A, B and C, matching the observed decrease in relaxing amplitude and increase of oscillating amplitude seen as $T_c$ is approached from below.

In conclusion, the single frequency~\musr~response and changes in the relaxing and oscillating amplitudes below $T_c=318$~K are best captured by the local fields at site A. 

\begin{table}
\begin{tabular}{c|c|c|c}
       Site & $|\bm{B}|$(mT) & $\nu$(MHz) & $\bm{\hat{B}}$ ~~\\
       \hline
         A & $1470$& $199$&($-0.002$, $-0.999$, $-0.054$) ~\\
         B & $730$& $99$&($0.002$, $-0.674$, $-0.739$)~\\
         C & $420$& $57$&($0.01$, $-0.999$, $-0.04$)~\\
         D & $60$& $8$&($-0.05$, $0.14$, $-0.99$)~\\
\end{tabular}
\caption{Dipole fields at the muon sites for the $a$-$b$ plane AFM structure in~\cmg~(for moments along the $b$ direction).}
\label{tab:Ce_AFM_fields}
\end{table}

\subsection{\pmg}

In~\pmg~our muon site calculations identify five classes of muon sites [see Tab.~\ref{tab:sym_muonsite_Pr}].
Sites A, B, C and D are equivalent to those realised in~\cmg, with only small differences in the distortion of the muon local environment. For example, site B in~\pmg~slightly perturbs both nearest Mn ions towards the muon ($<0.2$~\AA), while in~\cmg~one of the ions was slightly pushed away ($<0.2$~\AA). In contrast, site E is not realised in~\cmg~and sits at the centre of a pair of Pr ions separated by $\approx4$~\AA, perturbing them by less than $0.05$~\AA.
Based on their relative energies, we suggest it is not favourable for sites D and E to form in~\pmg.

\begin{table}
\begin{tabular}{c|c|c|c}
       Site & Coordinates & Relative energy (eV) & Wyckoff Site~~\\
       \hline
         A & (0, 0, 0.2) & 0.0 &$4e$~\\
         B & (0.14, 0.12, 0.18) & 0.15 &$32o$~\\
         C & (1/2, 0, 0.1) & 0.52 &$8g$~\\
         D & (0.43, 0.31, 0) & 0.83 &$16l$~\\
         E & (1/2, 0, 0) & 2.56 &$4c$~\\
         
\end{tabular}
\caption{Location, relative energies and symmetry of candidate muon sites calculated for~\pmg.}
\label{tab:sym_muonsite_Pr}
\end{table}

\begin{figure*}
  \centering
  \includegraphics[width=0.9\linewidth]{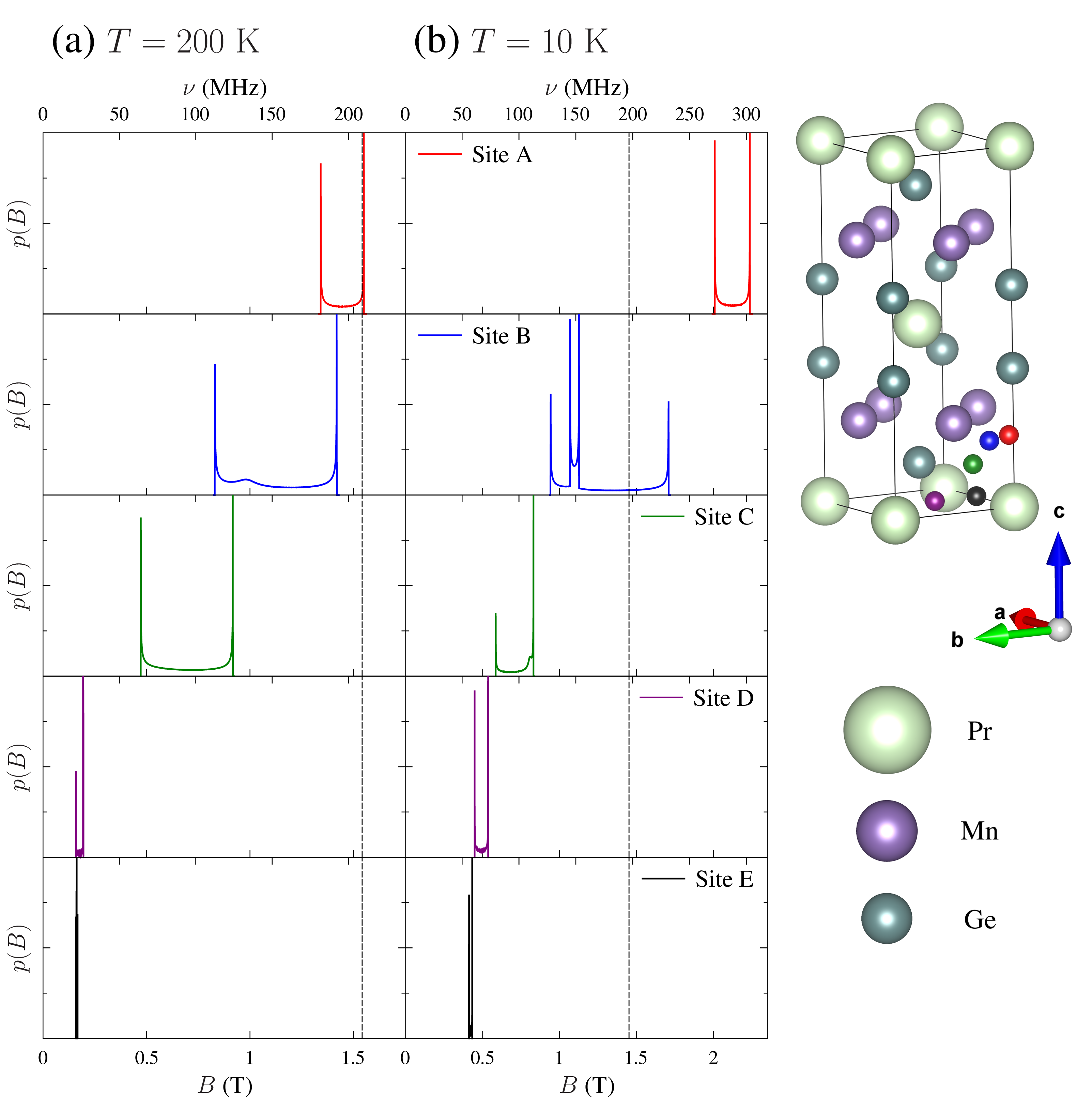}
  \caption{Dipolar field distributions calculated at each of the candidate muon sites in~\pmg~for the reported magnetic structures~\cite{Wang2014PrNeutrons} at (a) $200$~K and (b) $10$~K, where both the Mn and Pr moments are ordered. The dotted lines indicate the value of the local field from our~\musr~measurements, which corresponds to precession frequencies of $209.0(4)$~MHz at $200$~K and $196.8(4)$~MHz at $10$~K. Additionally, a visualisation of the locations of the muon sites in the~\pmg~unit cell is presented, where the colours of the spheres representing the muons corresponds to that of the field distributions.}
  \label{fig:Pr_field_dist}
\end{figure*}

Firstly, considering the I$_{c}$ phase at $200$~K (which is above any reported Pr ordering), the magnetic structure reported using neutron scattering is characterised by $m_\mathrm{Mn}=2.90(4)\mu_\mathrm{B}$, $\alpha=53.6(8)^\circ$ and $q_z=0.214(1)$, with the helical component of the Mn spins AFM aligned within the $a$-$b$ plane~\cite{Wang2014PrNeutrons}. 
The resulting field distribution for the five candidate muon sites [Fig.~\ref{fig:Pr_field_dist}(a)] show that site A has a field distribution with an average field of $1450$~mT ($197$~MHz), which is slightly smaller than the experimental precession frequency of $209.0(4)$~MHz. 
Site B has a field distribution average of $1100$~mT ($149$~MHz), while sites C, D and E all significantly underestimate the local field at the muon site. 
Our dipole field calculations for~\pmg~do not include the hyperfine contribution to the local field at the muon site: hyperfine fields of $90$~mT and $440$~mT are needed at sites A and B for the calculated average fields to agree with the experimentally observed frequency. 

As temperature decreases, the Mn moment size and propagation vector increase ($m_\mathrm{Mn}=3.23(7)\mu_\mathrm{B}$ and $q_z=0.272(2)$ at $10$~K), and the moments aligned closer to the $c$ axis ($\alpha=51.0(9)^\circ$). 
This is accompanied by FM aligning of the Pr spins along $c$ with moment $m_\mathrm{Pr}=1.53(3)\mu_\mathrm{B}$~\cite{welter1995neutrons,Wang2014PrNeutrons}.
Experimentally, we observed a decrease in the local field at the muon site of $90$~mT ($12.2$~MHz), but our calculations find an increase in average local field of $680$~mT ($91$~MHz) and $160$~mT ($21$~MHz) at sites A and B respectively. This increase in local field is due to the larger Mn moment at lower temperatures, as increasing the Pr moment size in our calculations reduces the average local field.
Moreover, increasing $m_\mathrm{Pr}$ to the largest physically reasonable value of $3.58\mu_\mathrm{B}$ (which corresponds to a free Pr$^{3+}$ ion~\cite{blundell2003magnetism}) still results in an increase of local field.
Finally, as was the case in~\cmg, the field distribution widths for site A agree much better with the single~\musr~frequency than site B in both the F$_\mathrm{Nd}$ and $I_c$ phases [Fig.~\ref{fig:Pr_field_dist}].

\begin{table*}
\begin{tabular}{c|c|c|c|c|c|c}
       Muon & \multicolumn{3}{c} {C phase} \vline& \multicolumn{3}{c} {AFM phase}~~\\
       Site & $|\bm{B}|$(mT) & $\nu$(MHz) & $\bm{\hat{B}}$& $|\bm{B}|$(mT) & $\nu$(MHz) & $\bm{\hat{B}}$  ~~\\
       \hline
         A & $1680$& $227$&($0.02$, $-0.91$, $-0.42$)& $1390$&$189$&($0.0$, $-0.998$, $-0.058$) ~\\
         B & $1200$& $163$&($0.17$, $-0.52$, $-0.84$)& $810$& $110$&($0.14$, $-0.82$, $-0.56$)~\\
         C & $610$& $82$&($-0.12$, $-0.99$, $0.02$)& $450$&$61$&($0.05$, $-0.79$, $-0.61$)~\\
         D & $120$& $17$&($-0.120$, $-0.005$, $-0.993$)& $40$& $5$&($-0.03$, $0.23$, $-0.97$)~\\
         E & $20$& $3$&($-0.001$, $-0.624$, $0.781$)& $20$& $3$&($-0.002$, $-0.5$, $-0.866$)~\\
\end{tabular}
\caption{Dipole fields at the muon sites for the canted (C) and the $a$-$b$ plane antiferromagnet (AFM) in~\pmg~(for moments along the $b$ direction).}
\label{tab:Pr_AFM_fields}
\end{table*}

As $T$ increases above $200$~K, the moment size and propagation vector of the Mn ion ordering decrease until a reported transition at $T_{c}=280$~K to a canted AFM.
The magnetic structure at $300$~K is defined by $m_\mathrm{Mn}=2.44(5)\mu_\mathrm{B}$ and a canting angle of $\alpha=58.4(6)^\circ$, with the component along the $c$ axis FM aligned and the components in the $a$-$b$ plane AFM aligned with neighbouring Mn ions~\cite{Wang2014PrNeutrons}. 
At $331$~K the moments then cant to $90^\circ$, forming an AFM phase constrained within the $a$-$b$ plane, with Mn moment $m_\mathrm{Mn}=1.86(3)\mu_\mathrm{B}$ at $350$~K~\cite{Wang2014PrNeutrons}.
While local fields at site A are mostly independent of the orientation of moments in the $a$-$b$ plane, this is not the case for site B (as would be expected for a lower symmetry site).
Therefore, we report the local fields along the $b$ axis [Tab.~\ref{tab:Pr_AFM_fields}], as they provide the best match to experiment for site B.
In the AFM phase the largest local field components at sites A and B are within the $a$-$b$ plane [Tab.~\ref{tab:Pr_AFM_fields}], agreeing with the large oscillating component present in the~\musr~above $T_c=331$~K. 
However, the local field magnitude at sites A and B are significantly larger than the observed frequency $\nu(T=350)=67.1(3)$~MHz. 
Then, as temperature decreases and the magnetic structure undergoes a canting transition, a larger component of local field is now orientated along the $c$ axis for sites A and B. 
Now the local field magnitude at site B is similar to the~\musr~frequency [$\nu(T=300~\mathrm{K}=138.9(6)$~MHz], but the field at site A is too large by $\approx660$~mT.
Finally, as the structure becomes incommensurate below $T_\mathrm{icm}\approx280$~K, the largest components of the local field distributions are along $c$ for both sites A and B, reflecting the relative sizes of the amplitudes seen in the~\musr~measurements.

In conclusion, while~\musr~measurements of the commensurate magnetic order seen above $T_\mathrm{icm}\approx280$~K in~\pmg~agree best with the calculated local fields at site B, only the local field distributions at site A agree with the fast single-frequency response seen below $T_\mathrm{icm}$. 
Therefore, on balance, this suggests that site A is the realised muon site in~\pmg. 
\subsection{\nmg}
We identify four classes of candidate muon sites in~\nmg, the same as those seen in~\cmg~with small differences in the distortion of the local muon environment.
 
\begin{table}
\begin{tabular}{c|c|c|c}
       Site & Coordinates & Relative energy (eV) & Wyckoff Site ~~\\
       \hline
         A & (0, 0, 0.2) & 0.50 &$4e$~\\
         B & (0.1, 0.1, 0.18) & 0.0 &$32o$~\\
         C & (1/2, 0, 0.1) & 0.45 &$8g$~\\
        D & (0.46, 0.22, 0) & 0.57 &$16l$~\\
\end{tabular}
\caption{Location, relative energies and symmetry of candidate muon sites calculated for~\nmg.}
\label{tab:sym_muonsite_Nd}
\end{table}

\begin{figure*}
  \centering
 \includegraphics[width=0.9\linewidth]{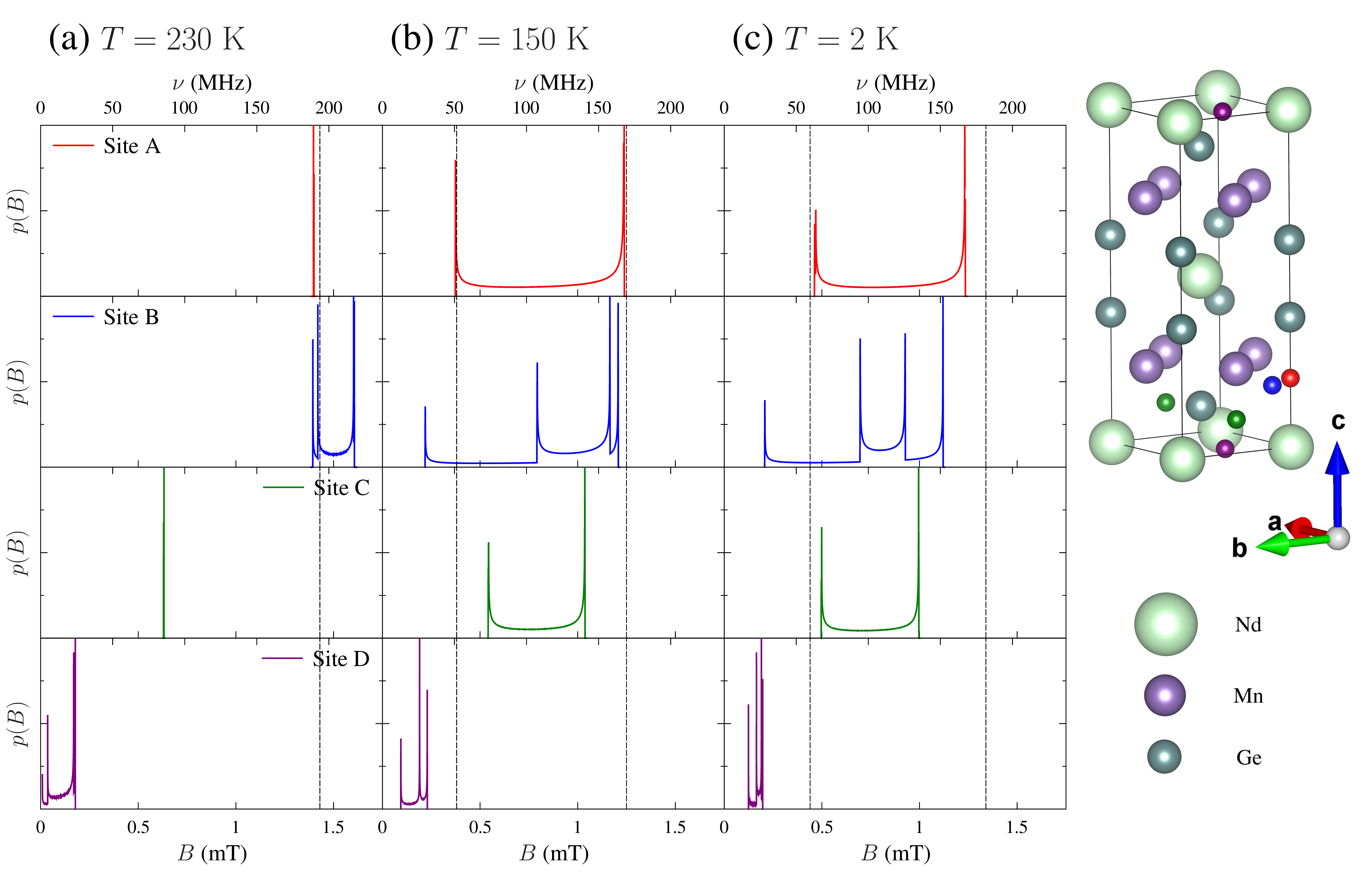}
  \caption{Dipolar field distributions calculated at each of the candidate muon sites in~\nmg~for the reported magnetic structures~\cite{welter1995neutrons,xu2023Ndtopological} at (a) $230$~K, (b) $150$~K, where the helical axis has reorientated along $b$, and at (c) $2$~K, where both the Mn and Nd moments are ordered. The dotted lines indicate the value of the local field from our~\musr~measurements, which corresponds to precession frequencies of $193.9(5)$~MHz at $230$~K, $170.0(8)$~MHz and $51.4(8)$~MHz at $150$~K, and $181.8(7)$~MHz and $60.1(7)$~MHz at $2$~K. A visualisation of the locations of the muon sites in the~\nmg~unit cell is presented, where the colours of the spheres representing the muons corresponds to the associated field distribution.}
  \label{fig:Nd_field_dist}
\end{figure*}

Firstly, we consider magnetic phases below the spin reorientation transition, $T_\mathrm{SR}=215$~K.
In the I$_{ab}$ phase, the reported magnetic structure at $150$~K is characterised by $m_\mathrm{Mn}=2.7\mu_\mathrm{B}$, $\alpha=58^\circ$, and  
$q_z=0.183(2)$, with the axis of the conical arrangement of Mn moments orientated within the $a$-$b$ plane~\cite{welter1995neutrons}.
The local field distributions calculated for sites A and C [Fig.~\ref{fig:Nd_field_dist}(b)] both have two peaks in field, matching the two-frequency response in the~\musr~measurements. 
For site A the peaks in field distribution, at $370$~mT and $1230$~mT, match the experimental local fields, $380(10)$~mT and $1250(10)$~mT, particularly well.
Above the spin reorientation transition, the switch to a one-frequency~\musr~response is seen in the field distributions at sites A and C [Fig.~\ref{fig:Nd_field_dist}(a)], but the local field calculated at site C is significantly smaller than the experimental value.
Finally, the component of the local field distribution along the $c$ axis is negligible in the I$_{ab}$ phase for site A, but not for site C, so the loss of the slow relaxing term in the~\musr~is only explained by the local field distribution at site A.

Below the Nd ordering temperature at $T_\mathrm{Nd}=21$~K, the magnetic structure at $2$~K is characterised by $m_\mathrm{Mn}=2.7\mu_\mathrm{B}$, $\alpha=56^\circ$, 
$q_z=0.227(1)$ and $m_\mathrm{Nd}=2.35(24)\mu_\mathrm{B}$, with the Nd moments order FM along the conical axis of the Mn structure~\cite{welter1995neutrons}.
This is similar to the low temperature change in magnetic order reported in~\pmg, but with no increase in $m_\mathrm{Mn}$.
A key feature of this transition in the~\musr~is that it leads to an increase in the smaller precession frequency equivalent to a field of $60$~mT.
Considering the field distribution at site A for a conical axis along the $a$ and $b$ axes, the lower peak in field is seen to decrease by $50$~mT along $a$, but increases by $90$~mT along $b$.
Meanwhile, for sites B and C the lower peak in field decreases by $10$~mT and $40$~mT with the conical axis along $b$.
Therefore, the orientation of the conical axis along $b$ leads to a local field distribution at site A that captures the increase of the smaller precession frequency, $\nu_2$, in the~\musr~measurements.

Above $T_\mathrm{SR}=215$~K,~\nmg~undergoes a similar set of magnetic transitions as~\pmg.
The I$_c$ phase (a conical helix of Mn moments with axis along $c$) occupies a narrow region in the phase diagram, between $215$~K and $240$~K. 
The magnetic structure in this phase has not been refined (apart from a propagation vector of $q_z=0.091$ at $230$~K) , so we have used the moment size and canting angle from the I$_{ab}$ phase ($m_\mathrm{Mn}=2.7\mu_\mathrm{B}$ and $\alpha=58^\circ$) in our dipole field calculations~\cite{welter1995neutrons,xu2023Ndtopological}. 
\nmg~then undergoes a transition to a canted AFM at $T_\mathrm{icm}=240$~K, which at $295$~K is characterised by $m_\mathrm{Mn}=1.8\mu_\mathrm{B}$ with a canting angle of $\alpha=58^\circ$ from the $c$ axis~\cite{welter1995neutrons}.
Finally, above $T_c=335$~K, the Mn moments form an AFM structure in the $a$-$b$ plane, with a moment size of $1.45(6)\mu_\mathrm{B}$ at $360$~K~\cite{Morellon1997_Nd}.

In the AFM phase the local field at sites A and C is orientated almost entirely within the $a$-$b$ plane [Tab.~\ref{tab:Nd_AFM_fields}], agreeing with the absence of a slow relaxing term in the~\musr. 
However, the local field at site A is almost double the field that would give the reported~\musr~frequency of $62.1(2)$~MHz at $360$~K.  
Below $T_c=335$~K, an increase in size of the $c$ component of the local field is seen for sites A, B and C, matching the reappearance of the slow relaxing term in the~\musr.
However, while the precession frequencies corresponding to the calculated local fields at sites A and B are similar to the experimental frequency of $146(1)$~MHz at $300$~K, site C has a local field that is almost three times too small.
Finally, in the I$_c$ phase [Fig.~\ref{fig:Nd_field_dist}(a)] sites A and B have local field distributions with peaks similar to the experimental~\musr~frequency, but only the orientation of the local field distribution at site A would result in a similar relaxing amplitude to the C phase.

In conclusion, the calculated local field distributions at site A provide the best fit to our~\musr~measurements.  
Moreover, the change in the smaller precession frequency, $\nu_2$, at the Nd ordering transition is only seen when the magnetic structures in the F$_\mathrm{Nd}$ and I$_{ab}$ phases are orientated along the $b$ axis.

\begin{table*}
\begin{tabular}{c|c|c|c|c|c|c}
       Muon & \multicolumn{3}{c} {C phase} \vline& \multicolumn{3}{c} {AFM phase}~~\\
       Site & $|\bm{B}|$(mT) & $\nu$(MHz) & $\bm{\hat{B}}$& $|\bm{B}|$(mT) & $\nu$(MHz) & $\bm{\hat{B}}$  ~~\\
       \hline
         A & $930$& $126$&($0.0$, $-0.85$, $-0.53$)& $740$&$101$&($0.0$, $-0.999$, $0.003$) ~\\
         B & $1010$& $137$&($0.15$, $-0.66$, $-0.74$)& $860$& $117$&($-0.01$, $-0.91$, $-0.41$)~\\
         C & $410$& $55$&($-0.002$, $-0.793$, $0.609$)& $310$&$41$&($-0.003$, $-0.999$, $0.0$)~\\
         D & $40$& $6$&($0.0$, $0.0$, $-1.0$)& $50$& $7$&($0.0$, $0.0$, $-1.0$)~\\
\end{tabular}
\caption{Dipole fields at the muon sites for the canted (C) and the $a$-$b$ plane antiferromagnet (AFM) in~\nmg~(for moments along $b$).}
\label{tab:Nd_AFM_fields}
\end{table*}

\section{Crystal Field Calculations}

The magnetism of rare-earth ions with partially filled $4f$ shells is understood by assuming that spin orbit interaction dominates over the crystal field splitting, due to the small spatial extent of the $4f$ orbitals and large size of the rare-earth nuclei. 
This couples the orbital ($\mathbf{L}$) and spin ($\mathbf{S}$) angular momenta, resulting in multiplets of energy levels that are described by a term symbol, $^{2S+1}L_{J}$, where $\mathbf{J}~(=\mathbf{L}+\mathbf{S})$ is the total angular momentum. The individual energy levels in these multiplets are labelled as eigenstates of total angular momentum, $\ket{m_J}$, where $-J\leq m_J\leq J$.
Therefore, in contrast to $3d$ transition-metals, the crystalline electric fields act as small perturbation on this $J$ multiplet, with the resulting energy level splitting dependent on the local symmetry of the rare-earth ion.
Due to this the associated crystal field Hamiltonian (for a single $J$ level) can be expressed as 
\begin{equation}
H_\mathrm{CEF}=\sum^{2l}_{k=0}\sum^k_{q=-k}B^q_kO^q_k,
\end{equation}
where $l=3$ for $f$ electrons, $B^q_k$ are crystal field parameters and $O^q_k=O^q_k(\mathbf{J})$ are the Stevens operator equivalents~\cite{Stevens1952_CFoperators,Hutchings1964_pointchargemodel}.

For~\remgall~the rare-earth ions sit at the $2a$ Wyckoff site with tetragonal symmetry, giving
\begin{equation}
    H_\mathrm{CEF}= B^0_2O^0_2 + B^0_4O^0_4 + B^4_4O^4_4 + B^0_6O^0_6 + B^4_6O^4_6
    \label{eq:tetragonalCFHamiltonian}
\end{equation}
as the resulting crystal field Hamiltonian~\cite{Fischer1987_tetragonalCF}.
For the Ce$^{3+}$ ion with ground state $\mathrm{^2F_{5/2}}$, the sixth order terms in Eq.~\ref{eq:tetragonalCFHamiltonian} are zero~\cite{Fischer1987_tetragonalCF} (as $J\leq5/2$). 
Diagonalising the resulting Hamiltonian into a $\ket{m_J}$ eigenbasis yields $\alpha\ket{\pm\frac{5}{2}}-\beta\ket{\mp\frac{3}{2}}$, $\beta\ket{\pm\frac{5}{2}}+\alpha\ket{\mp\frac{3}{2}}$ and $\ket{\pm\frac{1}{2}}$, with $\alpha=0.155$ and $\beta=0.988$, in the ground state multiplet.
Calculating the energies of these states using a point charge model~\cite{Hutchings1964_pointchargemodel} we find that $\ket{\pm\frac{1}{2}}$ forms a doublet CEF ground state, matching the ground state identified for~\cmg~in Ref.~\cite{MestnikFilho2007_electronicGS}, and we find an energy gap of $\Delta_\mathrm{ CEF}=68$~meV to the first excited doublet. 
For the Pr$^{3+}$ ion with a ground state $\mathrm{^3H_4}$, a similar procedure identifies a singlet CEF ground state $\ket{m_J}=\ket{0}$. An energy gap of $\Delta_\mathrm{ CEF}=9.7$~meV is calculated to the first excited doublet, $\alpha\ket{\pm1}+\beta\ket{\mp3}$ with $\alpha=0.99$ and $\beta=0.139$. %[Fig.~\ref{fig:RMn2Ge2_EL}].  
Finally the Nd$^{3+}$ ion with a ground state $\mathrm{^4I_{9/2}}$, has a doublet CEF ground state  $\alpha\ket{\pm\frac{1}{2}}+\beta\ket{\mp\frac{7}{2}}+\gamma\ket{\pm\frac{9}{2}}$ with $\alpha=0.985,\beta=0.156$ and $\gamma=0.074$. The first excited doublet, $\alpha\ket{\pm\frac{3}{2}}+\beta\ket{\mp\frac{5}{2}}$ with $\alpha=0.914$ and $\beta=0.407$, is $\Delta_\mathrm{ CEF}=1.7$~meV above the ground state. 

In relation to our~\musr~measurements, these energy level splitting become important for~\pmg.
This is because Pr$^{3+}$ has an even number of $4f$ electrons, so is a non-Kramers ion (i.e. doublets are not symmetry protected)~\cite{Kramers1930_thm,Klein1952_Kramers,blundell2003magnetism}.   
Therefore, a change in local symmetry of the Pr$^{3+}$ ion due to the muon implantation process could lead to the splitting of the first excited doublet in~\pmg; this effect is predicted by calculations on the distorted unit cell of~\pmg~using a point charge model.
Such a splitting alters the magnetism of the muonated material from that of the pristine material, which is reported in~\musr~measurements on other materials containing a Pr$^{3+}$ ion~\cite{Tashma1997_PrKramers,Foronda2015_PrKramers}.

\bibliography{bib}